\documentclass{aa}  
\usepackage{graphicx}
\usepackage{txfonts}
\usepackage{xcolor}
\usepackage{siunitx} 
\usepackage{hyperref}
\hypersetup{
  colorlinks   = true, 
  urlcolor     = blue,
  linkcolor    = blue, 
  citecolor   = blue 
}

\usepackage{gensymb}
\usepackage{amssymb}
\usepackage{amsmath}
\usepackage{pifont}

\usepackage{rotating}
\usepackage{longtable}
\usepackage{multirow}
\usepackage{booktabs}
\usepackage[makeroom]{cancel}
\usepackage[version=3]{mhchem}
\begin{document} 
\title{Spirals and Vertical Motions in the Planet-Forming Disk around HD\,100546}
\subtitle{A multi-line study of its gas kinematics}
\titlerunning{Spirals and Vertical Motions in HD\,100456}
\authorrunning{Wölfer et al.}
\author{Lisa W\"olfer\inst{1}\fnmsep\inst{2}, 
          Andr\'es F. Izquierdo\inst{2}\fnmsep\inst{3}\fnmsep\inst{4},
          Alice Booth\inst{2}\fnmsep\inst{5},
          Stefano Facchini\inst{6},
          Richard Teague\inst{1},
          Ewine F. van Dishoeck\inst{2}\fnmsep\inst{7},
          Teresa Paneque-Carre\~no\inst{2}\fnmsep\inst{4}\fnmsep\inst{8},
          Bill Dent\inst{9}
          }
\institute{Department of Earth, Atmospheric, and Planetary Sciences, Massachusetts Institute of Technology, Cambridge, MA 02139, USA
\and Leiden Observatory, Leiden University, P.O. Box 9513, NL-2300 RA Leiden, The Netherlands
\and Department of Astronomy, University of Florida, Gainesville, FL 32611, USA
\and European Southern Observatory, Karl-Schwarzschild-Str. 2, D-85748 Garching bei M\"unchen, Germany
\and Center for Astrophysics | Harvard \& Smithsonian, 60 Garden St., Cambridge, MA 02138, USA
\and Dipartimento di Fisica, Universit\`a degli Studi di Milano, Via Celoria 16, 20133 Milano, Italy
\and Max-Planck-Institut f\"ur extraterrestrische Physik, Gie\ss enbachstr. 1 , 85748 Garching bei M\"unchen, Germany
\and Department of Astronomy, University of Michigan, 323 West Hall, 1085 S. University Avenue, Ann Arbor, MI 48109, USA
\and Joint ALMA Observatory, Avenida Alonso de Córdova 3107, Vitacura 7630355, Santiago, Chile
   \\ e-mail: \href{mailto:lwoelfer@mit.edu}{lwoelfer@mit.edu}
             }
   \date{Received ; accepted}
%
%
  \abstract
   {The disk around the Herbig star HD\,100546 represents a particularly interesting target to study dynamical planet--disk interactions as various features have been observed in both the dust and gas that provide direct and indirect evidence for ongoing planet formation.}
   {In this work, we aim to characterize the gas kinematics of five molecular CO emission lines ($^{12}$CO 7-6, $^{12}$CO 3-2, $^{12}$CO 2-1, $^{13}$CO 2-1, C$^{18}$O 2-1), observed with ALMA in HD\,100546, to reveal deviations from Keplerian rotation as well as substructures in the peak intensity and line width.}
   {For our analysis, we fit the molecular intensity channels with the \textsc{discminer} package to model the line profiles, and extract observables such as centroid velocity, peak intensity, and line width. Aside from fitting the full cube, we also conduct runs where the blue- and redshifted sides are modeled separately to search for possible asymmetries.}
   {Our analysis reveals prominent kinematical spiral features in all five lines on large scales of the disk and we reproduce their morphology with both a linear and logarithmic spiral. In $^{12}$CO 2-1, spirals are also seen in the peak intensity residuals, the line width residuals exhibit a prominent ring of enhanced line widths around $125-330$\,au. The models further show, that the emission from the redshifted side may originate from higher disk layers than that from the blueshifted side, with the asymmetry being especially pronounced for $^{12}$CO 7-6.}
   {The pitch angles of the spirals are consistent with those driven by an embedded companion inside of 50\,au and they suggest a dynamical mechanism rather than gravitational instabilities. We further find indications of a companion around 90--150\,au, where tentative dips are present in the radial profiles of the integrated intensity of $^{13}$CO and C$^{18}$O 2-1 and pressure minima are observed in the azimuthal velocities. For the first time, we also detect downward vertical flows in this region, which coincide with the observed dust gap. The asymmetry in the emission heights may be a result of infall from the disk's environment. Another explanation is provided by a warped inner disk, casting a shadow onto one side of the disk.}
   \keywords{accretion, accretion disks --
                protoplanetary disks --
                planet-disk interactions --
                submillimeter: planetary systems --
                stars: individual: HD\,100546
               }
   \maketitle
%
\section{Introduction}
In order to understand planet formation and the diversity of developed planetary systems such as our own, it is crucial to study the early stages when (proto-) planets are still embedded in their birthplaces -- the so-called protoplanetary or planet-forming disks. These disks are not static objects but they evolve and eventually disperse over time with the evolutionary processes not only shaping the disk's appearance but also influencing (and putting a limit on) the planet-formation mechanisms. Conversely, planets interact with their environment, impact disk evolution, and are expected to alter their host disk's structure, for example in terms of density, temperature, or velocity. Even though such young embedded planets may be hidden from the direct eye of our telescopes, their planet--disk interactions still create signatures which are observable depending on the planet's mass and location. 

In the last decade, high-angular-resolution observations of the dust and gas in protoplanetary disks with the Atacama Large Millimeter/submillimeter Array (ALMA; \citealp{ALMA2015}), the Spectro-Polarimetric High-contrast Exoplanet REsearch (SPHERE; \citealp{Sphere2019}), or the Gemini Planet Imager (GPI; \citealp{Macintosh2014}) have enabled scientists to search for such signposts of planet--disk interactions. These observations have revealed a variety of substructures such as gaps, cavities, rings, spiral arms, azimuthal asymmetries, and shadows to be ubiquitous in both the dust and the gas component (e.g., \citealp{Bae2022,Benisty2022}). To interpret the origin of these substructures, it is necessary to understand how frequent they are, to discern possible patterns they follow, and to search for differences or similarities between different star-disk system morphologies.   

While various mechanisms such as photoevaporation (e.g., \citealp{Owen2011,Picogna2019}), gravitational instabilities (e.g., \citealp{Kratter2016}), magnetorotational instabilities (e.g., \citealp{Flock2015,Flock2017,Riols2019}), zonal flows (e.g., \citealp{Uribe2015}) or compositional baroclinic instabilities (e.g., \citealp{Klahr2004}) have been invoked to explain the observations, at least some of the substructures are expected to be linked to the presence of (massive) planets \citep{Lin1979,Zhang2018}. In this context, the so-called transition disks represent a particularly interesting subgroup of young stellar objects (YSOs). These disks are marked by inner regions depleted in dust (and gas) (e.g., \citealp{Espaillat2014,Ercolano2017}) and were originally identified through a lack of infrared (IR) excess in their spectral energy distribution (SED; \citealp{Strom1989}). While they may represent an intermediate (transitioning) state between an optically thick (full) disk and disk dispersal, dynamical clearing by a massive companion represents an alternative explanation. At least some of the cavities -- and especially the very deep ones ( e.g., \citealp{Marel2016}) -- are expected to be the result of such processes rather than representing an evolutionary state. Transition disks, therefore, represent ideal laboratories to probe disk evolution as well as planet formation models and may enable us to catch planet formation in action. 

Ultimately, only a direct detection can confirm the link between the observed disk substructures and an embedded planet. However, the dense and opaque environment of these young planets makes such a task difficult and feasible only for the very massive, bright planets that are less affected by dust extinction \citep{Sanchis2020}. To date, the only robust detections of forming planets (of several $M_{\mathrm{J}}$) and their circumplanetary disks have been obtained for the PDS\,70 system \citep{Keppler2018,Haffert2019,Benisty2021}, and more recently for the WISPIT 2 system \citep{Capelleveen2025,Close2025}. 

The challenge of direct detections and our growing understanding of how planets interact with their host disk have triggered the development of other, indirect, detection techniques. One promising method is to study the velocity field of the rotating gas, observed through molecular line emission, to search for deviations from Keplerian rotation \citep{Perez2015,Pinte2022}. Identifying such variations in the kinematics can be used to probe the local pressure gradient and to characterize the shape of the perturbation. Several ALMA observations in recent years have reported both localized and extended kinematical deviations, which may be linked to the presence of planets: so-called kink-features are detected by \cite{Pinte2018,Pinte2019} in HD\,163296 and HD\,97048 and a Doppler flip is reported in the HD\,100546 disk by \cite{Casassus2019}. \cite{Teague2018a} and \cite{Teague2019a} study the rotation profile of HD\,163296, finding perturbations and significant meridional flows that point towards embedded planets. \cite{Izquierdo2022} apply the \textsc{discminer} tool \citep{Izquierdo2021a} -- a new channel-map-fitting package able to model the upper and lower disk surfaces simultaneously and to identify localized velocity perturbations in both radius and azimuth -- to the same disk and find strong indications for two embedded planets. Extended spiral structures are observed in the kinematical residuals of TW\,Hya (\citealp{Teague2019Spiral,Teague2022}), HD\,100453 (\citealp{Rosotti2020a}), HD\,135344B \citep{Casassus2021}, CQ\,Tau (\citealp{Woelfer2021}), HD\,163296 and MWC\,480 \citep{Teague2021}, J1604 \citep{Stadler2023}, and HD\,142527 by \cite{Garg2021}. Moreover, the exoALMA large program conducted a comprehensive kinematical study of 15 disk to search for signatures of young embedded planets \citep{exoalma1}, revealing complex kinematic patterns in most disks and signatures consistent with a planet in six disks \citep{exoalma10}  

Aside from the kinematics, it is also interesting to study the patterns in the peak intensity/brightness temperature and line widths, which are both shaped by the presence of companions and can be used to trace density substructures, vertical motions, or turbulence. Planets can create density waves that result in an increased surface density and thus higher opacity, moving the $\tau = 1$ layer to a higher altitude where the temperature is generally higher (\citealp{Phuong2020b,Phuong2020a}). This leads to spiral substructures in the gas brightness temperature. Thermal spiral features are reported in TW\,Hya by \cite{Teague2019Spiral} and in CQ\,Tau by \cite{Woelfer2021}. These observations may be explained by the models of \cite{Muley2021}, who investigate the temperature structure in planet-driven spiral arms. \cite{Woelfer2022} study the gas brightness temperature (and kinematics) of a sample of 36 transition disks, finding significant substructures such as spirals in eight sources. Around the orbit of a massive planet, turbulent motions are triggered which are expected to result in enhanced line widths. \cite{Izquierdo2022} indeed observe an enhancement of line widths around the gas gap at 88\,au in HD\,163296, which further supports their findings of a planet candidate at 94\,au. Enhanced line widths are also found in HD\,135344B by \cite{Casassus2021}. 

In this work, we present a multi-line study of CO gas emission in the transitional disk HD\,100546, one particularly interesting target to study planet--disk interactions. The young Herbig Be star, located at $\sim$110\,pc from the Earth \citep{Gaia2018}, is surrounded by a massive disk spanning several hundreds of au and showing various indications of ongoing planet formation, as described in the next section. For our analysis, we apply the \textsc{discminer} package presented in \cite{Izquierdo2021a} to archival B6 ($^{12}$CO 2-1, $^{13}$CO 2-1, C$^{18}$O 2-1), B7 ($^{12}$CO 3-2), and B10 ($^{12}$CO 7-6) data to search for perturbations in the velocities, intensities, and line widths (possibly linked to embedded planets) and to obtain the vertical structure of the disk. Probing different layers of the disk with various isotopologues or transitions of the same molecule (as done in this study) or different molecules (e.g., \citealp{Paneque2022}) is crucial to understand how the observed substructures vary and can help us to access their formation mechanism (e.g., \citealp{Pinte2018a, Law2021}). We describe the source and observations in \hyperref[sec:observations]{Sect.~\ref*{sec:observations}}. Our modeling approach is detailed in {Sect.~\ref*{sec:analysis}} and the results are presented in {Sect.~\ref*{sec:results}}. The latter are discussed in {Sect.~\ref*{sec:discussion}} and summarized in {Sect.~\ref*{sec:summary}}. 
\begin{figure*}
\centering
\includegraphics[width=0.75\textwidth]{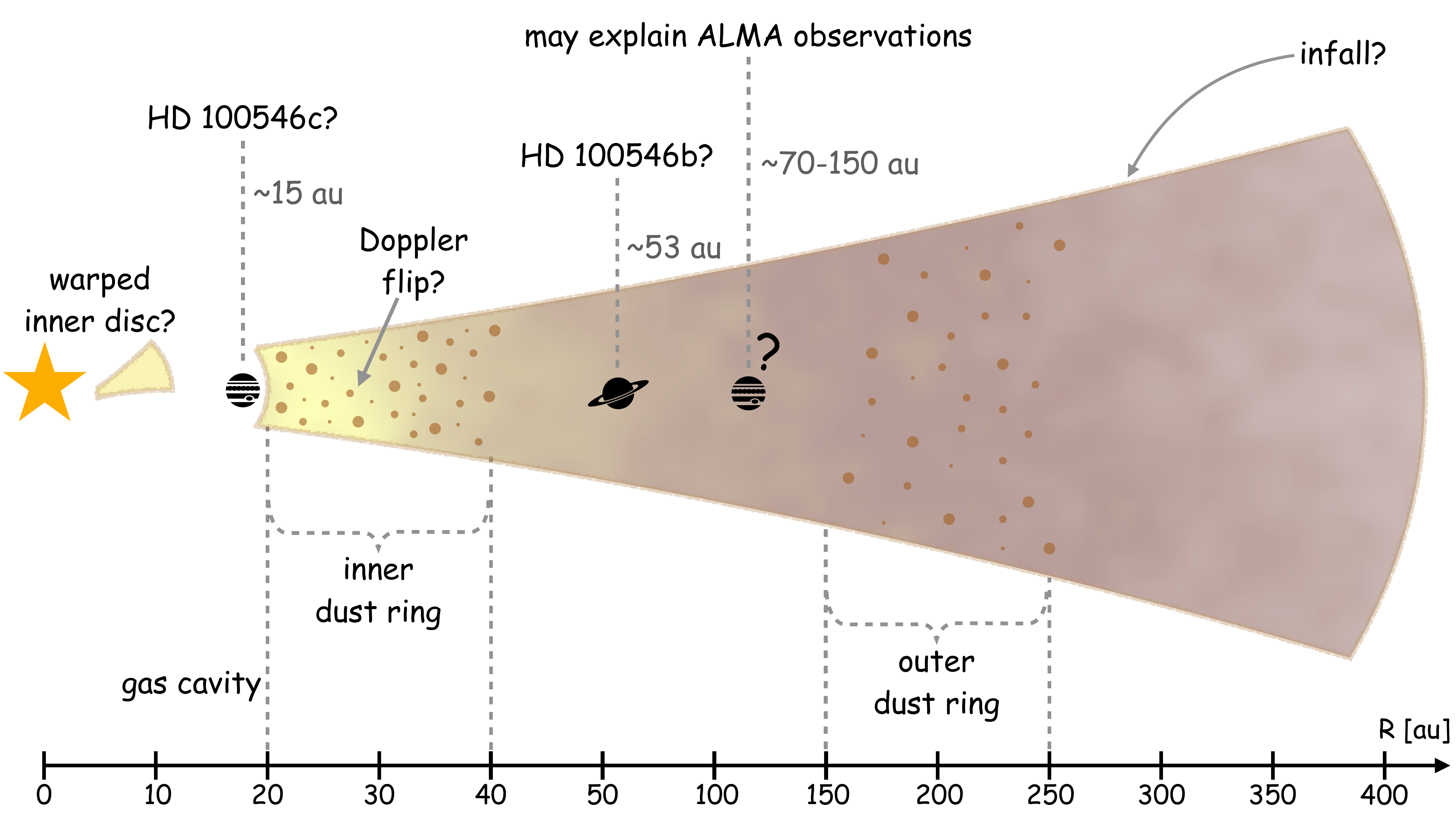}
\caption{Possible morphology of the HD\,100546 system. Substructures such as the dust rings have been observed, while the planet candidates have been proposed to explain the observations. The two planets suggested in the dust gap may refer to the same object for which the actual location is yet undetermined.}\label{fig:cartoon}
\end{figure*} 
\section{Observations}\label{sec:observations}
\subsection{The HD\,100546 system}
In \hyperref[fig:cartoon]{Fig.~\ref*{fig:cartoon}}, we show the possible morphology of the HD\,100546 system further described in the following. Observations of a compact source (in $L'$-Band) have been interpreted as direct evidence of an embedded companion in the outer disk at $\sim$ 53\,au \citep{Quanz2013,Quanz2015,Currie2015b}, also known as the giant planet candidate HD\,100546\,b. Subsequent $Ks$-Band observations, however, revealed a faint extended emission around the same location \citep{Boccaletti2013} and thus the planetary origin of the observed feature yet needs to be confirmed. Another giant planet, HD\,100546\,c, may be located at $\sim$14\,au, just inside the dust cavity. First postulated by \cite{Acke2006} and later identified as a point source in near-infrared scattered light data taken with GPI \citep{Currie2015b}, the detection of HD\,100546\,c has been debated in the literature (e.g., \citealp{Folette2017,Currie2017}). The presence of the planet is supported by studies on the time variations of ro-vibrational CO emission lines \citep{Brittain2009,Brittain2013,Brittain2014,Brittain2019}, tracing the inner edge of the disk. These observations showed an excess of CO P(26) line emission between 2003 and 2013 -- consistent with an orbiting planet -- that disappeared in 2017 as the possible planet moved behind the near side of the inner rim of the outer disk. Assuming a common origin, the feature observed in the scattered light is expected to disappear as well and recent observations with VLT/SPHERE indeed lack a detection \citep{Sissa2018}.
\begin{table*}
\centering 
\small
\caption{Observational properties of the ALMA data sets used in this work.}
\label{tab:dataOrig}
\begin{tabular}{lcccccccc}
\toprule
\toprule
Project ID & Date & Baselines & Int. time & Central freq. & Bandwidth & $\Delta \upsilon$ & Ang. res. & LAS \\
& & (m) & (min) & (GHz) & (MHz) & (km$\,$s$^{-1}$) & ($\arcsec$) & ($\arcsec$) \\
\midrule
2011.0.00863.S & 18 Nov 2012 & 21--375 & 12 & 345.798 & 469 & 0.211 & 0.57 & 6.66\\
Cycle 0 & \\[0.8\normalbaselineskip]
2015.1.00806.S & 02 Dec 2015 & 17--10800 & 30 & 345.796 & 469 & 0.212 & 0.035 & 0.471\\
Cycle 3 & \\[0.8\normalbaselineskip]
2016.1.00344.S & Oct 2016 - & 19--12200 & 35, 117 & 230.538 & & 0.158 & 0.183, 0.037 & 1.983, 0.745\\
Cycle 4 & Sep 2017 & & & 220.399 & 117 & \\
& & & & 219.560 & & \\[0.8\normalbaselineskip]
2018.1.00141.S & 14 Apr 2019 & 15--780 & 28 & 807.499 & 1875 & 0.181 & 0.165 & 2.007\\
Cycle 6 \\
\bottomrule
\end{tabular}
\end{table*}

Alongside the direct measurements, the HD\,100546 disk also exhibits a number of indirect evidence for planet--disk interactions. The scattered light images are marked by spiral structures (e.g., \citealp{Garufi2016,Folette2017}) and the millimeter continuum, well studied with ALMA, reveals a large inner cavity ($\lesssim$ 20\,au) with the dust being concentrated into two prominent rings between 20--40\,au and 150--250\,au (e.g., \citealp{Walsh2014,Pineda2019,Fedele2021}). Models show that these substructures are consistent with two giant planets orbiting at locations of 10--15\, au and 70--150\,au \citep{Pinilla2015,Fedele2021,Pyerin2021}. In addition to the dust, the gas disk of HD\,100546 has been studied through a range of molecular tracers. ALMA observations of the CO isotopologues are numerous and show that the disk is much more extended in the gas than in the dust (e.g., \citealp{Walsh2014, Miley2019}).  

The $^{12}$CO emission revealed non-Keplerian features in the kinematics that may be a signpost of ongoing planet formation: \cite{Walsh2017} find asymmetries that could be explained by a warped inner disk, which however stands in contrast to SPHERE observations of the disk by \cite{Garufi2016}, who suggest radial flows mediated by planets as an alternative explanation. A warp has also been proposed by \cite{Panic2010} to account for asymmetric line observations ($^{12}$CO $J=6-5$ and $J=3-2$) from APEX. As mentioned before, \cite{Casassus2019} detect a Doppler flip at $\sim$28\,au which is in agreement with the spiral wake created by a planet and coincides with a fine ridge in the millimeter continuum, suggesting a complex dynamical scenario. The Doppler flip is confirmed by \cite{Casassus2022}, with the blueshifted side of the flip however disappearing when vertical and radial flows are taken into account. The authors propose an embedded outflow, launched by a companion of several $M_{\mathrm{Earth}}$, as a possible explanation. This could explain the observations by \cite{Booth2018}, who detect non-Keplerian features in the blueshifted side of SO line emission that may be tracing a disk wind, a warped disk, or an accretion shock onto a CPD. Follow-up observations by \cite{Booth2022} support that the azimuthal SO asymmetry is indeed tracing an embedded planet. Another interpretation of the Doppler flip as being the kinematic counterpart of the NIR spiral is proposed by \cite{Norfolk2022}. \cite{Perez2020} analyze the wiggles seen in the velocity channels of $^{12}$CO, $^{13}$CO, and  C$^{18}$O. They find that the most pronounced wiggle in $^{12}$CO resembles the imprint of an embedded massive planet. The strength of the wiggles decreases for $^{13}$CO and C$^{18}$O, and thus with disk height, suggesting that the perturbed flow is associated with vertical motions in the disk surface. As shown by \cite{Izquierdo2021a}, such kink-like features can also be caused by a gap or density substructures rather than a planet and caution needs to be exercised in the interpretation of these perturbations. 

HD\,100546 is known to still be embedded in a faint envelope \citep{Grady2001,Ardila2007}, reaching out to about 1000\,au. Such an envelope may be linked to secondary accretion events due to infalling material onto the disk, which is often accompanied by arc-shaped structures as observed in HD\,100546 \citep{Dullemond2019}. Asymmetric accretion from an envelope can result in a tilt of the disk \citep{Thies2011}, the formation of vortices \citep{Bae2015}, or drive spirals due to accretion shocks, that propagate the disk \citep{Lesur2015,Hennebelle2017}. Spirals have been observed in HD\,100546 in both the optical \citep{Grady2001,Ardila2007} and the near-infrared \citep{Avenhaus2014,Sissa2018}.     

In this paper, we aim to shed some additional light onto the observed kinematical deviations by analyzing different CO tracers in data sets that cover both the inner and outer disk with the channel map fitting package \textsc{discminer}.
\subsection{ALMA observations}
Molecular line emission of the disk around HD\,100546 has been observed with ALMA in several frequency bands. In this paper, we make use of the data sets 2011.0.00863.S (PI: C. Walsh; \citealp{Walsh2014}), 2015.1.00806.S (PI: J. Pineda; \citealp{Pineda2019}), 2016.1.00344.S (PI: S. P{\'e}rez; \citealp{Perez2020}), and 2018.1.00141.S (PI: B. Dent) that cover CO isotopologues in Bands 6, 7, and 10. Their properties are summarized in \autoref{tab:dataOrig}. For all of the data sets, we started with the archival pipeline-calibrated data and performed further reduction using CASA version 5.7.0 \citep{McMullin2007}. 

The Band 6 data of the \ce{^{12}CO}, \ce{^{13}CO}, and \ce{C^{18}O} $J=2-1$ line were taken in two configurations during Cycle 4, an extended (C40-9) as well as a compact configuration (C40-6) with the same spectral setting. We self-calibrated and combined these data following the same procedure as described in \cite{Czekala2021}: we self-calibrated the short-baseline data first and then combined it with the calibrated long-baseline data, after ensuring that the data sets share a common phase centre. Afterwards, we applied further iterations of self-calibration, both in phase and amplitude.

For the combination of the Band 7 data, we took a slightly different approach as the data sets do not share a significant overlap in the spectral settings (aside from both targeting the \ce{^{12}CO} $J=3-2$ line). The long-baseline data taken during Cycle 3 have insufficient short baselines to properly recover spatial scales $> 1\farcs0$. To improve this, we combined the data with observations from Cycle 0. We note, that since these observations were taken in different cycles and taken with different settings, they are not optimal for data combination. Each data set was first self-calibrated (phase and amplitude) individually. The Cycle 0 data were initially pipeline calibrated in CASA version 3.4 and therefore we applied the CASA task \texttt{statwt} to these data to account for the change in the visibility weight initialization and calibration across different CASA versions $< 4.3$ \footnote{\url{https://casadocs.readthedocs.io/en/stable/notebooks/data_weights.html}}. The self-calibrated data sets were then combined after ensuring they share a common phase centre. For the Band 10 data, no self-calibration was performed due to a lack of signal-to-noise-ratio (SNR). 

The continuum was subtracted using the \texttt{uvcontsub} task, flagging channels that contained line emission. We then imaged the lines with \texttt{tCLEAN} for both the continuum-subtracted and non-subtracted data sets. In the cleaning process, we used a Briggs robust weighting of $+0.5$, the `multi-scale' deconvolver, and a Keplerian mask \footnote{\url{https://github.com/richteague/keplerian_mask}}. A slight \texttt{uv-taper} was applied in all Bands in order to improve the SNR in the images. The properties of the final images are summarized in \autoref{tab:dataInfo2}. As high spectral resolution is essential to study kinematics, we imaged the lines with the best possible spectral resolution. However, to ease comparison between the lines, we also made images for the same velocity resolution, given by the lowest resolution among the data sets (B7, 210\,m\,s{$^{-1}$}). We conducted our full analysis on both the image cubes described above and on image cubes with artificially reduced noise levels using the so-called JvM correction \citep{Jorsater1995,Czekala2021}. In the following, we only present the results for the non-corrected cubes, while the JvM-corrected data are included in \autoref{appendix:jvm}. Our results are consistent among both data sets. In Figs. \ref{fig:Channels12co32} and \ref{fig:Moments12co32}, we show a comparison between the final images of the individual and combined data sets for the Band 7 \ce{^{12}CO} $J=3-2$ line, where it becomes very clear that the inclusion of the shorter baseline data significantly improves the recovery of flux and image fidelity in the outer disk. 
\begin{table}[h!]
\centering
\small
\caption{Characteristics of the non-JvM corrected data analyzed in this work.}
\label{tab:dataInfo2}
\begin{tabular}{l cccc}
\toprule
\toprule
Line & $\Delta \upsilon$ & Beam & RMS & Line peak\\
& (km$\,$s$^{-1}$) & ($\arcsec$) & \multicolumn{2}{c}{(mJy$\,$beam$^{-1}$)} \\
\midrule
\rule{0pt}{10pt}$^{12}$CO 2-1 & 0.17 & 0.10x0.08 & 1.9 & 70.9\\
$^{13}$CO 2-1 & 0.17 & 0.10x0.08 & 1.9 & 30.6\\
C$^{18}$O 2-1 & 0.17 & 0.10x0.08 & 1.6 & 21.4\\
$^{12}$CO 3-2 & 0.21 & 0.12x0.09 & 8.8 & 215.8\\
$^{12}$CO 7-6 & 0.185 & 0.23x0.17 & 87.6 & 4207.5\\
\bottomrule
\end{tabular}
\end{table} 
\begin{figure*}
\centering
\includegraphics[width=1.0\textwidth]{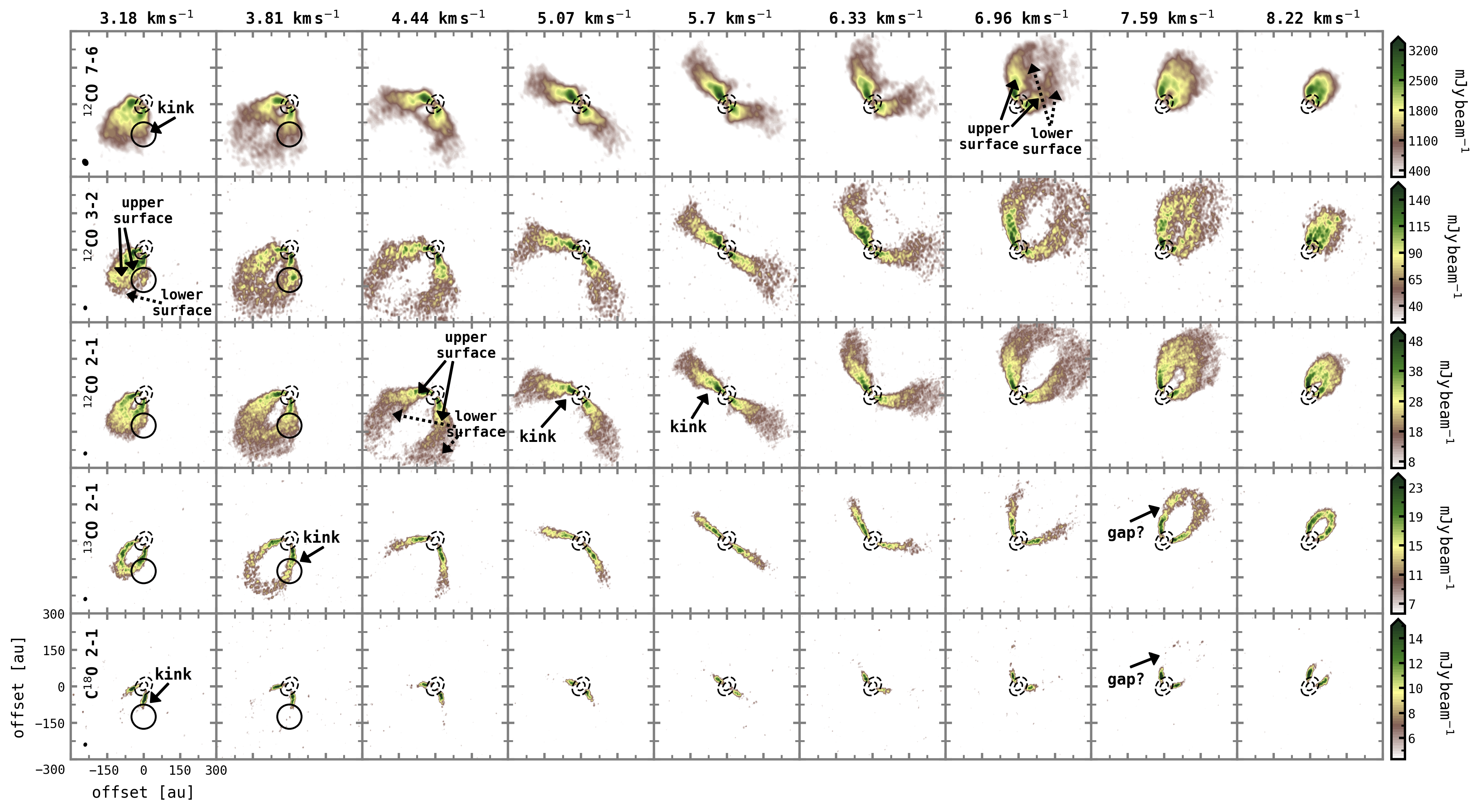}
\caption{Examples of the intensity channels of the five CO lines, shown in steps of three from the central channel and for a velocity resolution of 210\,m\,s$^{-1}$. The inner ring of the millimeter continuum is overlaid as dashed contours. The upper and lower disk surface are annotated in some channels as well as a potential gap and velocity kinks. A localized feature is seen in some of the channels within the solid circle. The beam of the observation is indicated in the bottom left corner of the first column panels. The channel maps are masked below 3\,$\sigma$.}\label{fig:Channels}
\end{figure*}
\begin{figure*}
\centering
\includegraphics[width=1\textwidth]{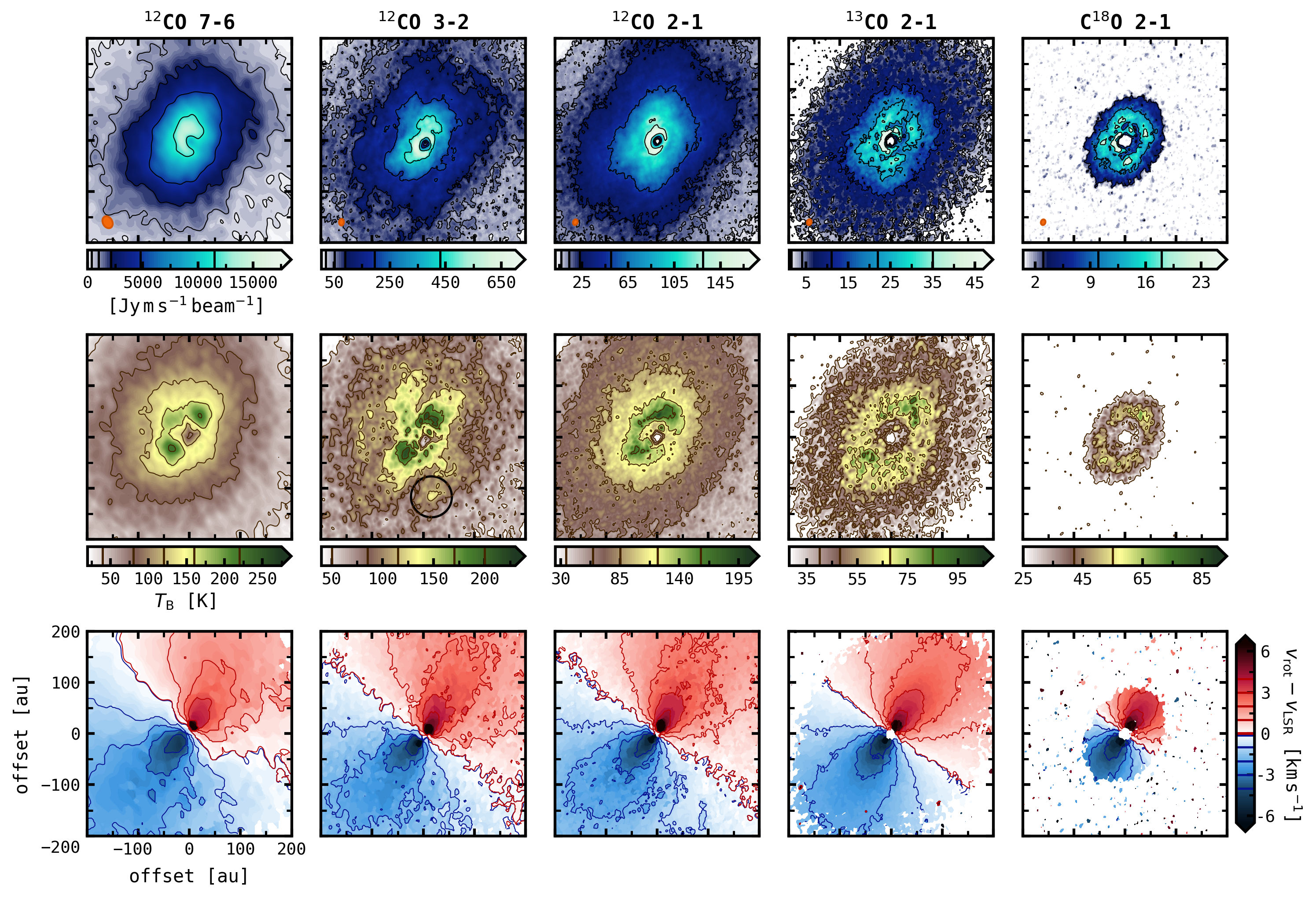}
\caption{Integrated intensity (top), peak intensity (middle), and line-of-sight velocity (bottom) of the five CO lines studied here. A localized feature is seen in $^{12}$CO 3-2 around 125\,au, marked by the solid circle. Some contours are overlaid and their levels are indicated in the color bars. The beam of the observation is shown in the bottom left corner of the first-row panels. The integrated intensity maps are masked at 3$\sigma$, the peak intensity maps at 5$\sigma$.}\label{fig:Moments}
\end{figure*}
\begin{figure}
\centering 
\hspace{-0.8cm}\includegraphics[width=0.42\textwidth]{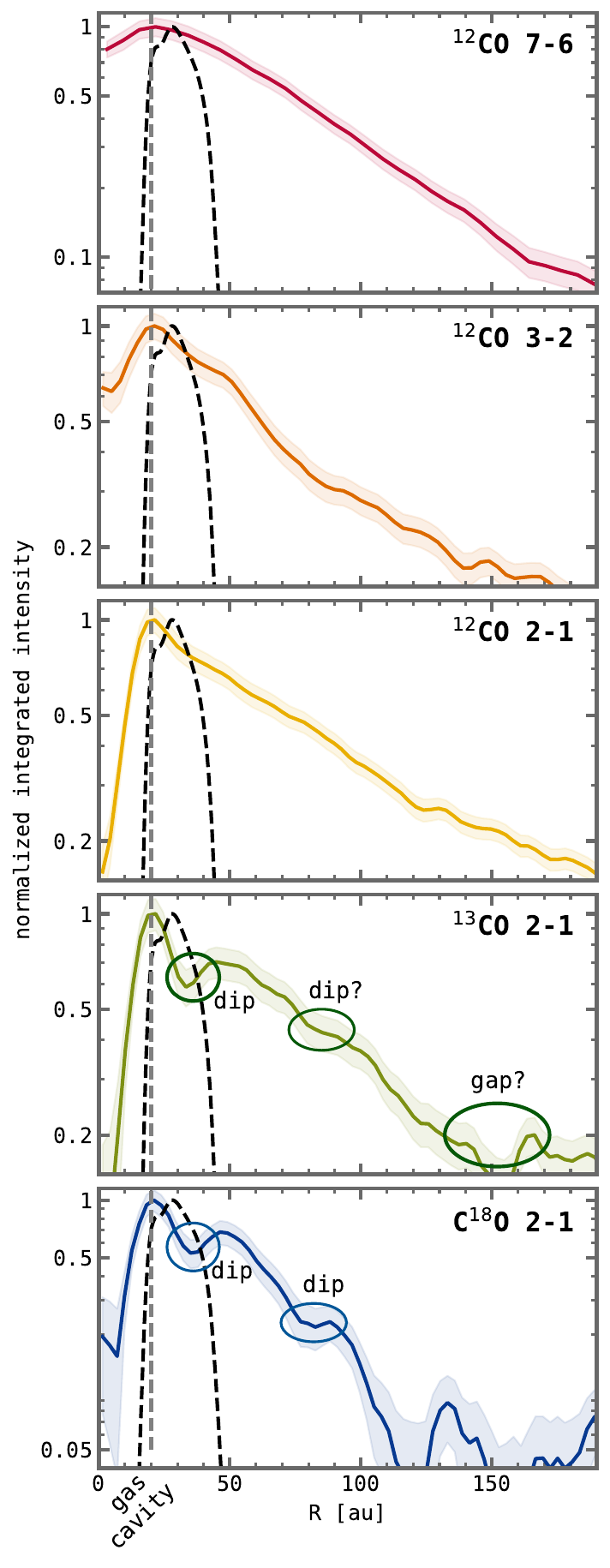}
\caption{Azimuthally averaged and normalized radial integrated intensity profiles for the different CO lines. A radial profile of the Band 6 millimeter continuum is included as black dashed lines. The vertical grey dashed line marks the location inside which the gas emission rapidly drops (gas cavity). Some features of the profiles are annotated.}\label{fig:RadialProfiles}
\end{figure}
\subsection{Observational results}\label{sec:resultsobs}
In \hyperref[fig:Channels]{Fig.~\ref*{fig:Channels}}, we present the continuum-subtracted intensity channels of the five CO lines in steps of three from the central channel and for the same channel spacing of 210\,m\,s$^{-1}$. The inner continuum ring is overlaid as dashed contours, showing that the gas disk is much more extended than the dust disk. As seen in the first three rows of \hyperref[fig:Channels]{Fig.~\ref*{fig:Channels}}, the $^{12}$CO data appear elevated with the lower surface of the disk being at least partly visible. The other two isotopologues, shown in the bottom two rows, on the other hand, appear rather flat and the lower surface can not be distinguished. This suggests that the emission from these molecules comes, as generally expected \citep{Law2021}, from lower heights in the disk located closer to the midplane. For these rarer isotopologues, line wing emission may be missing due to sensitivity effects. 

In the first two channels of the $^{12}$CO 3-2 data, a bright spot is visible at a separation of $\sim 100$\,au to the south of the central star, highlighted by the solid black circle in Figs. \ref{fig:Channels} and \ref{fig:Channels12co32}. As those data represent a combination of ALMA Cycle 0 and Cycle 3 data, both time variations and a difference in the observational setups may lead to imaging artifacts. It could further be the result of a projection effect. However, looking at the other lines, a similar, yet much weaker spot is present in the $^{12}$CO 2-1 data and a wiggle or break can be discerned in the corresponding $^{12}$CO 7-6 and $^{13}$CO 2-1 channels around this area. Altogether, this points towards the localized feature being real, but higher sensitivity data including both the long and the short baselines in the same observational setup are needed to support this claim. Continuum subtraction can be ruled out as a cause for the substructure as the continuum ring is located much closer to the star. A comparison of the combined data with the long- and short-baseline data is included in Appendix \ref{appendix:chans12co32}. 

The channels of $^{12}$CO 7-6 appear to be very asymmetric: the iso-velocity curves seem to be twisted in the centre with the two sides bending in opposite directions. Expected to come from the highest layers of the disk, this may result from interactions with material surrounding the disk but could also point towards misaligned disk regions (e.g., \citealp{Walsh2017, Facchini2018}). Furthermore, the northeast side of the disk appears to be brighter than the southwest side of the disk. The $^{13}$CO and C$^{18}$O data show tentative signs of an outer intensity gap in the disk, becoming most prominent for the 7.59\,km\,s$^{-1}$ channel, but higher sensitivity data are needed to confirm its presence. As mentioned in previous work \citep{Perez2020}, several of the velocity channels are marked by wiggles or kink-like features in their profile, some of which are annotated in \hyperref[fig:Channels]{Fig.~\ref*{fig:Channels}}. 

In \hyperref[fig:Moments]{Fig.~\ref*{fig:Moments}} we show the integrated and peak intensity maps alongside the kinematics for all five lines. These maps were computed with the \textsc{bettermoments} package \citep{bettermoments} and in case of the peak intensity converted to units of Kelvin with the Planck law. While the intensity maps were obtained with the standard moment 0 and 8 implementations, respectively, the kinematics were derived with the quadratic method. In this approach, a quadratic function is fitted to the brightest pixel in the spectrum as well as the two neighboring pixels to find the centroid of the line in pixel coordinates. We note, that for optically thick lines, continuum subtraction can remove part of the line emission and artificially lower the peak intensity, as part of the underlying continuum is absorbed (e.g., \citealp{Weaver2018, Rosotti2021, Bosman2021}). While we did not apply any masking in the computation of the peak intensity maps, a Keplerian mask was included in the computation of the integrated intensity, and for the kinematics we selected a masking for regions below a certain SNR to reduce the noise at the disk's edge (between 4--5).

As seen in the intensity maps, an inner gas cavity (< 20\,au) is present in all isotopologues and lines. At the same location as in the channel maps, a localized, however weak spot is seen in the peak intensity of $^{12}$CO 3-2 (black circle in Figs. \ref{fig:Moments} and \ref{fig:Moments12co32}), the other lines are lacking a similar feature at this location. In general, the $^{12}$CO 3-2 intensities appear to be very asymmetric, with a cross-like -- or possibly spiral-like -- structure extending towards the north and south. Again, caution should be exercised with regard to these substructures, as artifacts may have been introduced through the combination of data sets. In \hyperref[fig:Moments12co32]{Fig.~\ref*{fig:Moments12co32}} we compare the intensity and kinematical maps for the different $^{12}$CO 3-2 data sets. Even though the long-baseline observation may be lacking flux in some regions, the cross-like structure is also present in the short-baseline data, pointing towards a real substructure and suggesting that no strong artifacts have been introduced in the imaging process. With three years having passed between the two observations, some time variations can, however, not be ruled out. It is puzzling that no similar structures are found in the $^{12}$CO 2-1 and $^{12}$CO 7-6 data. While the lack of features in $^{12}$CO 2-1 may be explained by temperature effects with the substructures becoming visible only at a certain height above the midplane, a similar pattern would at least be expected in $^{12}$CO 7-6, which traces even higher disk layers. Interaction with material from the environment of the disk could, however, wash out substructures. Moreover, $^{12}$CO 7-6 may be lacking flux in the outer disk regions. To get to the bottom of these peculiar substructures, it is crucial to obtain higher sensitivity data with a similar observational setup in the future.  

On first look, the kinematics overall suggest an only mildly elevated disk: despite an inclination of $\sim$ 40\,$\degree$ only slight contributions from the back side of the disk are visible in $^{12}$CO and no contributions are seen in $^{13}$CO and C$^{18}$O. In the $^{12}$CO 7-6 kinematics, the central velocity seems to be off by a few 100\,m\,s$^{-1}$ from the systemic velocity of the source at 5.7\,km\,s$^{-1}$. This is also seen in the according channel map in \hyperref[fig:Channels]{Fig.~\ref*{fig:Channels}} which seems `S-shaped' as the south wing is bent upwards compared to the north wing. The central channel of the other lines on the other hand appears more symmetric. Assuming that $^{12}$CO 7-6 emission comes from a higher disk layer, the disk's surface may be distorted due to interactions with material from the environment. We analyze the kinematics in further detail in \hyperref[sec:analysis]{Sec.~\ref*{sec:analysis}}

\autoref{fig:RadialProfiles} presents normalized azimuthally averaged radial profiles of the integrated intensity for the lines of this analysis and the continuum emission. These profiles were obtained with the \textsc{gofish} package \citep{Gofish} by shifting and stacking all line spectra to increase the SNR. This approach is based on the assumption that the disk is symmetric and Keplerian, which does not necessarily hold. Therefore, we compared the profiles to those obtained from simply azimuthally averaging the moment 0 maps, yielding similar results. By default, the widths of the annuli in the \textsc{gofish} package are given as one-fourth of the beam major axis. For the deprojection, we assume a disk geometry as obtained from the models presented in \hyperref[sec:analysis]{Sec.~\ref*{sec:analysis}} and \autoref{tab:fits}. The gas intensity significantly drops inside of $\sim$ 20\,au for all lines, thus just inside the dust cavity. We note the location of the drop as the gas cavity, but this represents an upper limit. To get a better estimate of the actual gas cavity, a more thorough analysis of the temperature structure across the cavity such as in \cite{Leemker2022} or a kinematical approach such as in \cite{Bosman2021} is required. In addition to the inner cavity, a prominent drop of emission is seen in $^{13}$CO and C$^{18}$O at $\sim$ 35\,au. This intensity drop coincides with the continuum emission and thus may be caused by continuum absorption rather than representing a gas gap. Another intensity drop is observed in C$^{18}$O around 85\,au (and potentially $^{13}$CO) and in $^{13}$CO around 150\,au, the C$^{18}$O emission is too noisy around this location. This dip corresponds to the gap tentatively detected in the channel maps. Continuum absorption is unlikely responsible for this substructure as it peaks around 200\,au, but (particularly for  C$^{18}$O) higher sensitivity data are needed to quantify a gas gap at these radii.
\section{Analysis}\label{sec:analysis}
\subsection{Model setup}
We use the \textsc{discminer} code introduced in \cite{Izquierdo2021a} to model the intensity channels (see \hyperref[fig:Channels]{Fig.~\ref*{fig:Channels}}) of the HD\,100546 disk. In the following, we outline the basic modeling approach but refer the reader to \cite{Izquierdo2021a} for the full details of the package. In the \textsc{discminer}, parametric prescriptions are adopted to reproduce the intensity channel maps from molecular emission. The parameters used in this work are briefly summarized as follows (see \autoref{tab:ParamMiner}). 
\begin{table}
\centering
\small
\caption{Parameterization adopted for the \textsc{discminer} models in this work\tablefootmark{(a)}.}\label{tab:ParamMiner}
{\renewcommand{\arraystretch}{1.4}%
\begin{tabular}{ll}
\toprule
\toprule
Attribute &  Parameterization\\
\midrule
Orientation & $i$, PA, $x_{\mathrm{c}}, y_{\mathrm{c}}$ \\
Rotation velocity & $v_\mathrm{K} = \sqrt{G M_*/r^3} \cdot R$, $v_{\mathrm{sys}}$ \\
Emission surface & $z_{u,l} = z_{u0,l0}(R/D_0)^p$ $\cdot$ exp $[-(R/R_\mathrm{t})^q]$ \\
Peak intensity & $I_{\mathrm{p}} = I_0 (R/D_0)^{p0} (z/D_0)^q$, $R < R_{\mathrm{break}}$\\
               & $I_{\mathrm{p}} = I_0 (R/D_0)^{p1} (z/D_0)^q$, $R > R_{\mathrm{break}}$ \\
& $I_{\mathrm{p}} = 0$, $R > R_{\mathrm{out}}$ \\
Line width & $L_{\mathrm{w}} = L_{\mathrm{w0}} (R/D_0)^p (z/D_0)^q$ \\
Line slope & $L_{\mathrm{s}} = L_{\mathrm{s0}} (R/D_0)^p$ \\
\bottomrule
\end{tabular}
\tablefoot{\tablefoottext{a}{$D_0$ represents a normalization constant, set to 100\,au, $R$ and $z$ are the disk cylindrical coordinates, and $r$ the spherical radius. $G$ is the gravitational constant, all other variables are left as free parameters.}}
}
\end{table} 
\begin{table*}
\small
\centering
\caption{Best-fit results of the modeling of the full cubes with the \textsc{discminer}.}\label{tab:fits}
\begin{tabular}{lccccccc}
\toprule
\toprule
Attribute &  Parameter & Unit & $^{12}$CO 7-6 & $^{12}$CO 3-2 & $^{12}$CO 2-1 & $^{13}$CO 2-1 & C$^{18}$O 2-1 \\
\midrule
\multirow{4}{*}{Orientation} & $i$ & $\degree$ & 39.30 & 44.29 & 44.40 & 43.03 & fixed\\
& PA & $\degree$ & 320.95 & 321.31 & 323.37 & 324.45 & fixed \\
& $x_{\mathrm{c}}$ & au & --0.44 & 4.86 & 1.45 & 0.36 & 1.52 \\
& $y_{\mathrm{c}}$ & au & --2.20 & --7.06 & --1.38 & --1.04 & 9.12 \\
\midrule
\multirow{2}{*}{Velocity} & $M_{*}$ & M$_{\odot}$ & 2.21 & 1.93 & 2.05 & 2.16 & 2.02 \\
& v$_{\mathrm{sys}}$ & km\,s$^{-1}$ & 5.38 & 5.65 & 5.65 & 5.64 & 5.57\\
\midrule
\multirow{4}{*}{Upper surface} & $z_{0}$ & au & 19.32 & 23.11 & 19.99 & 2.59 & 2.65 \\
& $p$ & - & 0.85 & 0.86 & 0.82 & 0.02 & 2.56 \\
& $q$ & - & 1.88 & 2.59 & 3.68 & 4.99 & 1.14 \\
& $R_{\mathrm{t}}$ & au & 292.86 & 306.15 & 329.45 & 212.03 & 165.52 \\
\midrule
\multirow{4}{*}{Lower surface} & $z_{0}$ & au & - & 13.37 & 25.98 & 43.08 & - \\
& $p$ & - & - & 1.05 & 1.74 & 0.98 & - \\
& $q$ & - & - & 1.07 & 0.98 & 2.48 & - \\
& $R_{\mathrm{t}}$ & au & - & 341.71 & 142.61 & 21.25 & - \\
\midrule
\multirow{6}{*}{Intensity} & $I_{0}$ & Jy\,px$^{-1}$ & 7.04 & 0.63 & 0.24 & 0.06 & 0.009 \\
& $p_0$ & - & --0.97 & 0.59 & --0.18 & 0.92 & 0.94 \\
& $p_1$ & - & --1.40 & --0.82 & --1.04 & --0.43 & --2.46 \\
& $q$ & - & 0.83 & 0.40 & 0.61 & 0.30 & --0.001 \\
& $R_{\mathrm{break}}$ & au & 54.98 & 42.87 & 55.15 & 58.62 & 64.27 \\
& $R_{\mathrm{out}}$ & au & 322.04 & 393.46 & 375.35 & 326.57 & 253.61 \\
\midrule
\multirow{3}{*}{Line width} & $L_{\mathrm{W}}$ & km\,s$^{-1}$ & 2.01 & 0.70 & 0.68 & 0.28 & 0.02 \\
& $p$ & - & --0.90 & --0.49 & --0.56 & --0.82 & --1.28 \\
& $q$ & - & 0.44 & --0.05 & 0.03 & --0.10 & --0.87 \\
\midrule
\multirow{2}{*}{Line slope} & $L_{\mathrm{S}}$ & - & 2.03 & 1.94 & 2.12 & 1.75 & 1.79 \\
& $p$ & - & 0.29 & 0.27 & 0.37 & 0.05 & --0.001 \\
\bottomrule
\end{tabular}
\end{table*}

To describe the orientation of the disk, we use the position angle PA, inclination $i$, and disk center ($x_{\mathrm{c}}, y_{\mathrm{c}}$). For the rotation velocity we adopt a Keplerian rotation, using the stellar mass $M_*$ and source systemic velocity $v_{\mathrm{sys}}$ to describe the background velocity. The upper and lower emission surfaces are controlled by the height $z$ above and below the disk midplane. Here we use a power law with an exponential tapering. The line peak intensity, line width, and line slope across the disk radial and vertical extent are parameterized with simple power laws of the disk cylindrical coordinates ($R, z$). For the peak intensity we select a combination of two power laws, describing the intensity profile inside and outside a radius $R_{\mathrm{break}}$. By doing so, we are able to account for a decrease in intensity inside the gas cavity. For radii larger than the parameter $R_{\mathrm{out}}$, the peak intensity is set to zero. 

Subsequently, the \textsc{discminer} is coupled with the Markov chain Monte Carlo (MCMC) random sampler \textsc{emcee} \citep{emcee} to find those model parameters that best recover the intensity of the observed data cube. The different physical and morphological properties of the disk are thus modeled simultaneously to give a comprehensive view of the gas substructures and kinematics. From the disk attributes, a model for the line profiles and data channels is generated. The modeled intensity $I_m$ is described by a generalized bell kernel 
\begin{equation}
I_{\mathrm{m}} (R, z, v_{\mathrm{ch}}) = I_{\mathrm{p}} \left(1+| \frac{v_{\mathrm{ch}}-v_{\mathrm{K^{l.o.s}}}}{L_{\mathrm{w}}}|^{2L_{\mathrm{s}}} \right)^{-1}  
\end{equation}
as a function of the disk cylindrical coordinates $(R, z)$ and velocity channel $v_{\mathrm{ch}}$. $I_{\mathrm{p}}$ is the peak intensity, $L_{\mathrm{w}}$ half the line width at half power, $L_{\mathrm{s}}$ the line slope, and $v_{\mathrm{K^{l.o.s}}}$ the Keplerian line-of-sight velocity. The vertical coordinate $z$ is determined by the height of both the upper and the lower disk surface, thus each property except for $L_s$ has two descriptions. The contribution of the upper and lower emitting surfaces are merged into a single-line profile by selecting the highest intensity between bell profiles computed for both surfaces independently in each velocity channel and pixel.
\begin{figure*}
\centering
\includegraphics[width=1.0\textwidth]{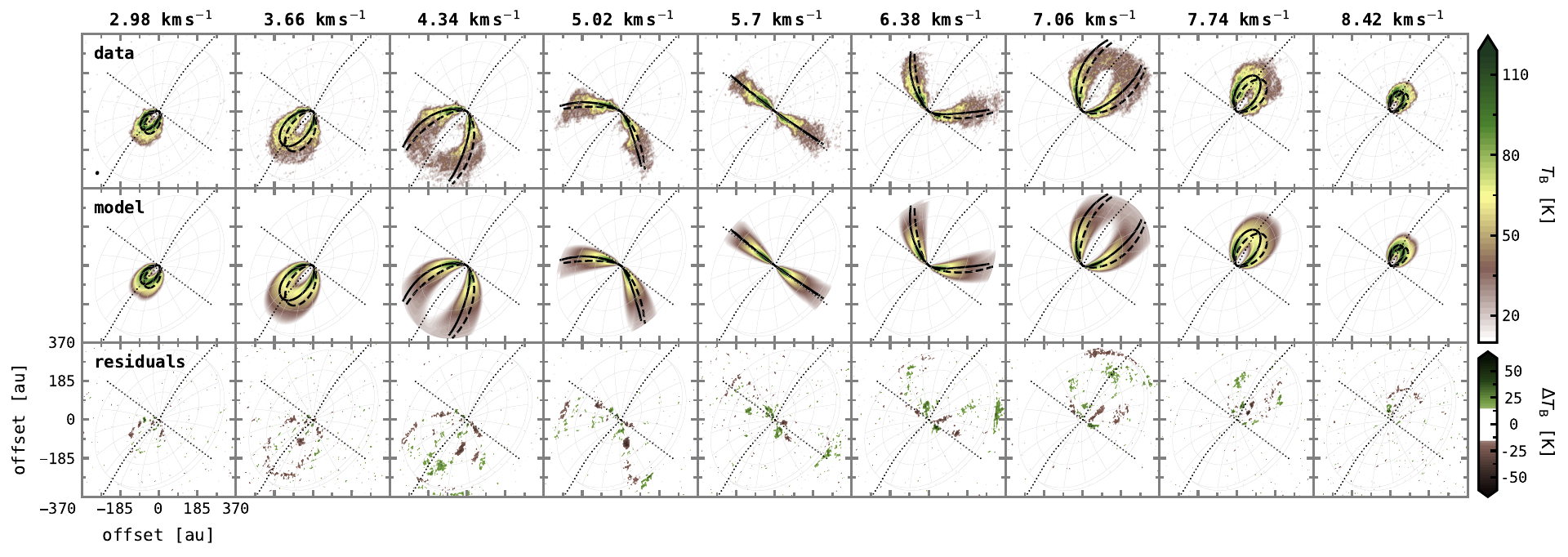}
\caption{Examples of the intensity channels of $^{12}$CO 2-1 (top), shown in steps of four around the central channel, alongside the best-fit model obtained with the \textsc{discminer} (middle) and the corresponding residuals (bottom). The iso-velocity curves of the upper and lower surfaces are overlaid as solid and dashed lines, respectively. The emitting surface and disk axes are plotted in the background as gray contours and dotted lines, respectively. The beam of the observation is indicated in the bottom left corner of the first panel.}\label{fig:ChannelsComp12co21}
\end{figure*}

To initialize the \textsc{emcee} sampler, we chose inclination, position angle, and stellar mass found by the models of \cite{Woelfer2022} and guessed the other parameters using the interactive prototyping tool of the \textsc{discminer}, which allows to compare the morphology of the model and data channel maps as well as line profiles by eye. The MCMC was then run with a varying number of walkers (depending on the number of fitting parameters) until convergence and the posterior distribution was sampled from the last 1000 steps. The noise of the data was taken into account in the fit: at each data pixel, the standard deviation of the residual intensities in line-free channels was calculated and adapted as a weighting factor for the likelihood function to be maximized by the sampler (see \citealp{Izquierdo2021a}). To ensure that the noise of the individual pixels was approximately independent of their neighboring pixels, we down-sampled the data with at least one beam while still ensuring that we had enough pixels left for the fit ($> 50$).

The HD\,100546 disk has previously been proposed to be warped in the center \citep{Walsh2017}. This could result in a temperature difference, potentially producing an elevation asymmetry, as it has been observed in CO emission of the Elias\,2-27 disk \citep{Paneque2021}. Moreover, the system is still surrounded by a diffuse envelope \citep{Grady2001} and the infall of material from that envelope onto the disk could also result in such an asymmetry. To check for asymmetries between the blue- and redshifted sides of the disk, we conducted additional runs where only half of the channels of the according side were modeled instead of using the full cube. In these runs we fixed the disk orientation, i.e. the inclination and position angle, to those of the full cube $^{12}$CO model, to assure that any potential asymmetries are independent of the disk's orientation. Similar, for C$^{18}$O we fixed the disk orientation (see \citealp{Izquierdo2025}) and switched off the lower surface in all runs since the channels appear to be very flat. Even though this is true also for $^{13}$CO, we ran models fitting for both surfaces and models where the lower surface was not included, yielding similar results. In the following, we will thus use the models where both surfaces were modeled. In the case of $^{12}$CO 7-6, only the models that accounted for the blue- and redshifted side separately converged, when the lower surface was included in the fit. To model the full cube, we therefore conducted additional runs with the lower surface switched off, which converged fine and are thus used in our following results. 
\begin{figure*}[h!]
\centering
\includegraphics[width=0.75\textwidth]{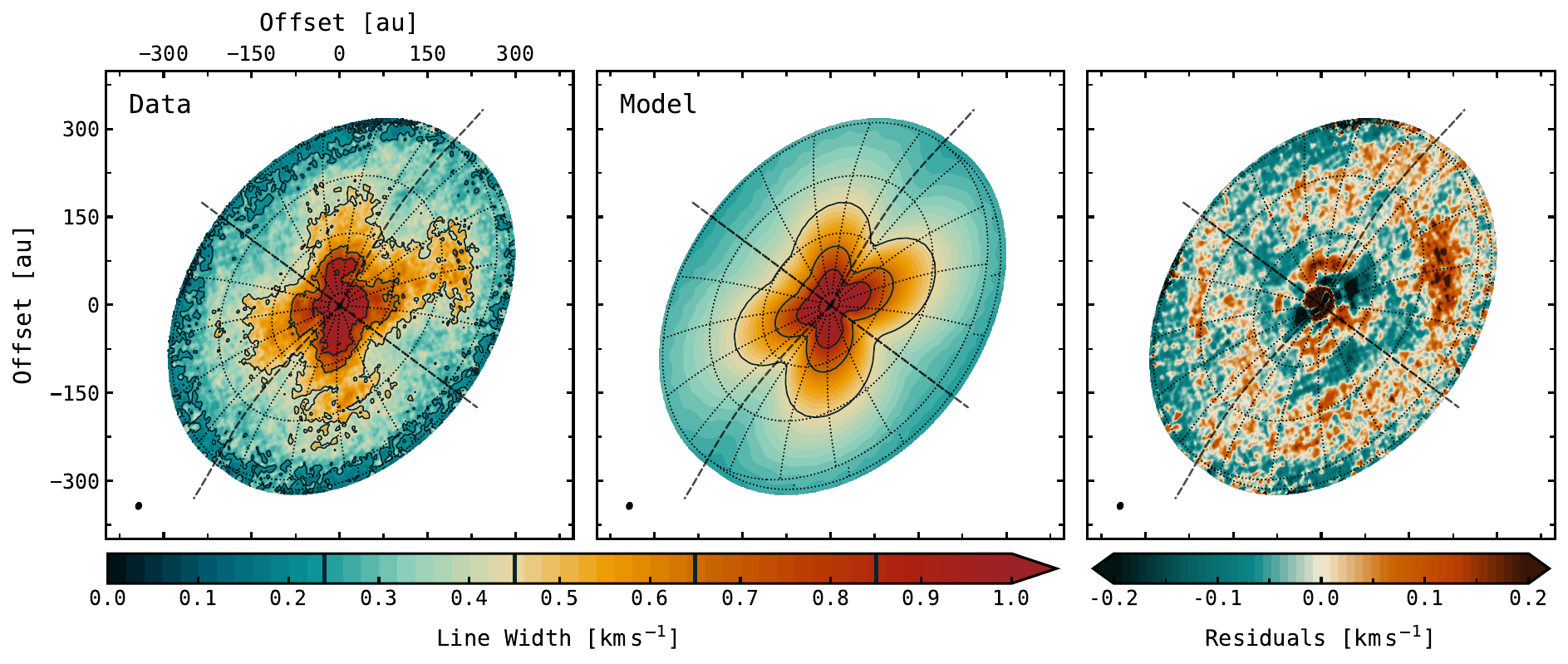}
\includegraphics[width=0.75\textwidth]{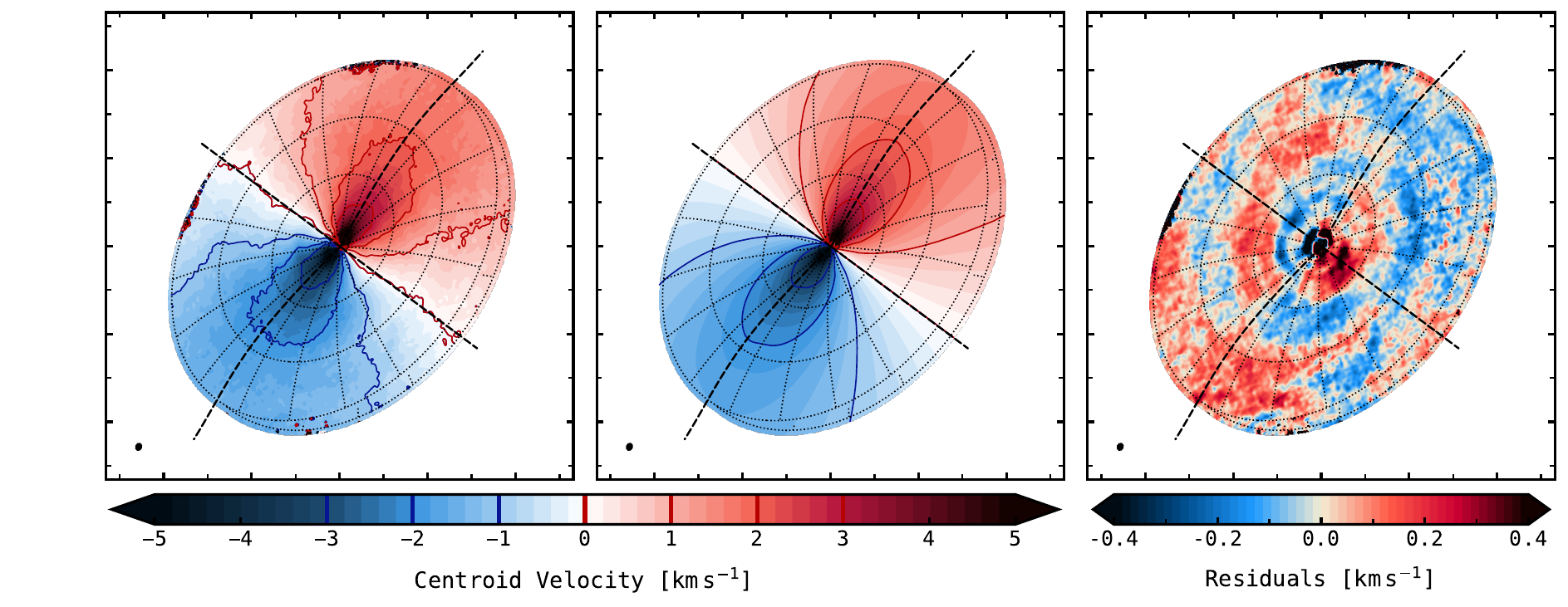}
\includegraphics[width=0.75\textwidth]{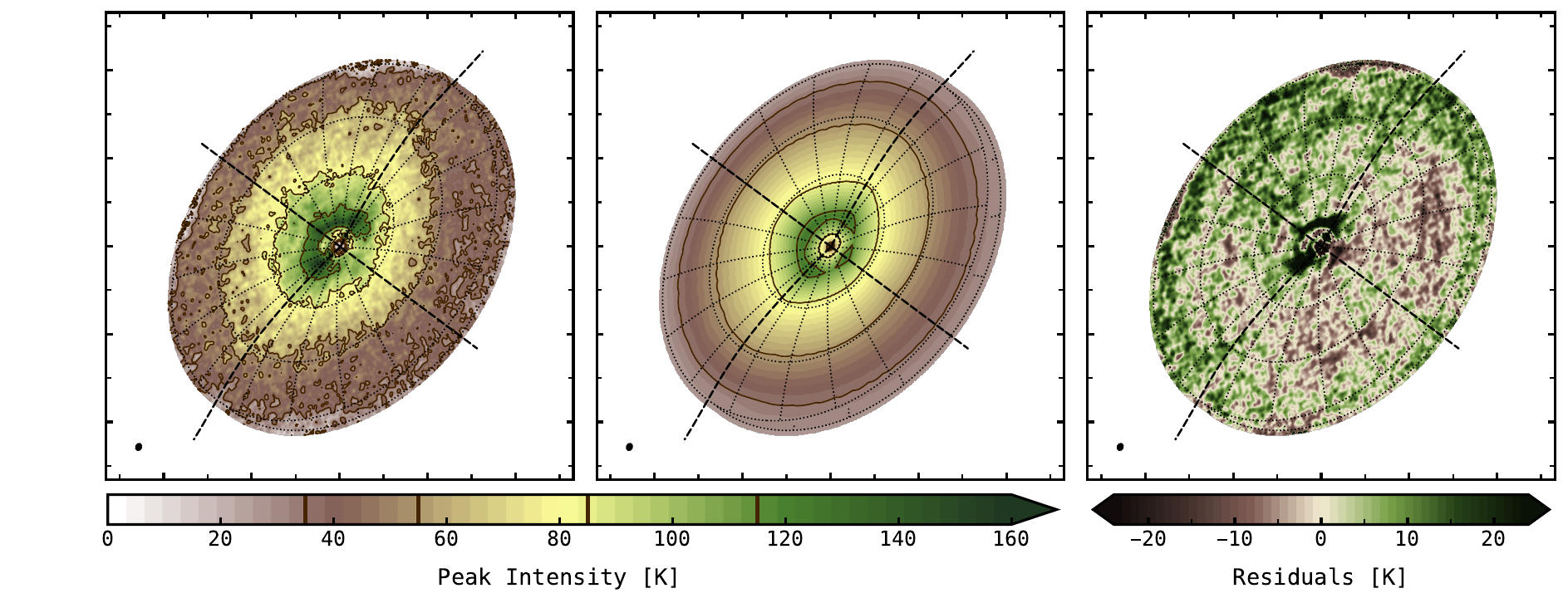}
\caption{Observables extracted from the modeled line intensity, shown for $^{12}$CO 2-1 as an example. Displayed are the line widths (top), centroid velocities (middle), and peak intensities (bottom) for the data and best-fit model. The residuals, showing the differences between the data and model, are included in the last column. Some contour levels, indicated in the color bars, are included to ease comparison. The emission surface and disk axes are plotted as dotted and dashed lines, respectively.}\label{fig:Observables}
\end{figure*}
\subsection{Fitting Results}
The best-fit parameters are summarized in \autoref{tab:fits} for the models fitting the full cube and in Tables \ref{tab:fitsblue} and \ref{tab:fitsred} for the models fitting the blue- and redshifted side separately. The retrieved orientation of the disk is very similar in all models, resulting in a mean inclination $i$ and position angle PA of $42.76 \pm 2.39\,\degree$ and $322.52 \pm 1.67\,\degree$, respectively. We note that the convention used for the position angle is that it is measured from the north to the red-shifted side of the disk. A slightly larger, but still small scatter is found for the stellar mass ($2.07 \pm 0.11\,M_{*}$), which is relatively sensitive to changes in the model. For the Band 6 and Band 7 data, the systemic velocity is very close to the expected value of 5.7\,km\,s$^{-1}$, however, for the $^{12}$CO 7-6 it is fitted as 5.4\,km\,s$^{-1}$. As mentioned before, this offset from the source velocity found in the literature (e.g., \citealp{Walsh2014,Walsh2017}) can be seen in the channel and rotation maps. The trend of decreasing line width from $^{12}$CO 7-6 to C$^{13}$O is picked up by the \textsc{discminer}. The radius $R_{\mathrm{break}}$,  where a switch in the peak intensity profiles is happening, generally increases towards C$^{18}$O (exception is $^{12}$CO 3-2). This can be explained by the inner gas cavity (< 20\,au) becoming more pronounced in the more optically thin lines. We note however, that $R_{\mathrm{break}}$ marks the point where the intensity is maximal and not the actual cavity radius, and other effects (such as temperature) influence its location. 
\subsubsection{Channel maps}
In \hyperref[fig:ChannelsComp12co21]{Fig.~\ref*{fig:ChannelsComp12co21}}, we present a comparison of the data and best-fit model channels of ${^{12}}$CO 2-1 alongside the residuals obtained from subtracting the model from the data. The channels are shown in steps of four around the central channel. The according plots of the other lines are included in Appendix \ref{appendix:chancomp}. The overall morphology of the data channels is well reproduced by the models, except for ${^{12}}$CO 7-6, where the lower surface could not be modeled properly. The emission surfaces obtained from the models are overlaid as gray contours in the individual panels. For all lines, their morphology matches that of a rather flat and not very flared disk: the back side of the disk is barely visible and the azimuthal contours do not show any clear bending. While this is not unusual for $^{13}$CO and C$^{18}$O, it is puzzling that we do not see a stronger manifestation of the vertical structure in $^{12}$CO. 

The surface of the disk is expected to become most visible in $^{12}$CO 7-6. In \hyperref[fig:ChannelsComp12co76both]{Fig.~\ref*{fig:ChannelsComp12co76both}}, we therefore plot the channels as a combination of the models fitting only half of the cube. While the emission surface on the blueshifted side is still showing a flat morphology, the redshifted side appears much more elevated and flared. A similar exercise for the other $^{12}$CO lines did not yield a significantly different emission surface compared to the one retrieved from the full cube. Another interesting point about the combined $^{12}$CO 7-6 plot is that it shows that the vertical structure itself can produce apparent wiggles in the channels, which are nicely retrieved by the model, especially for the blueshifted side.
\begin{figure*}[h!]
\centering
\includegraphics[width=\textwidth]{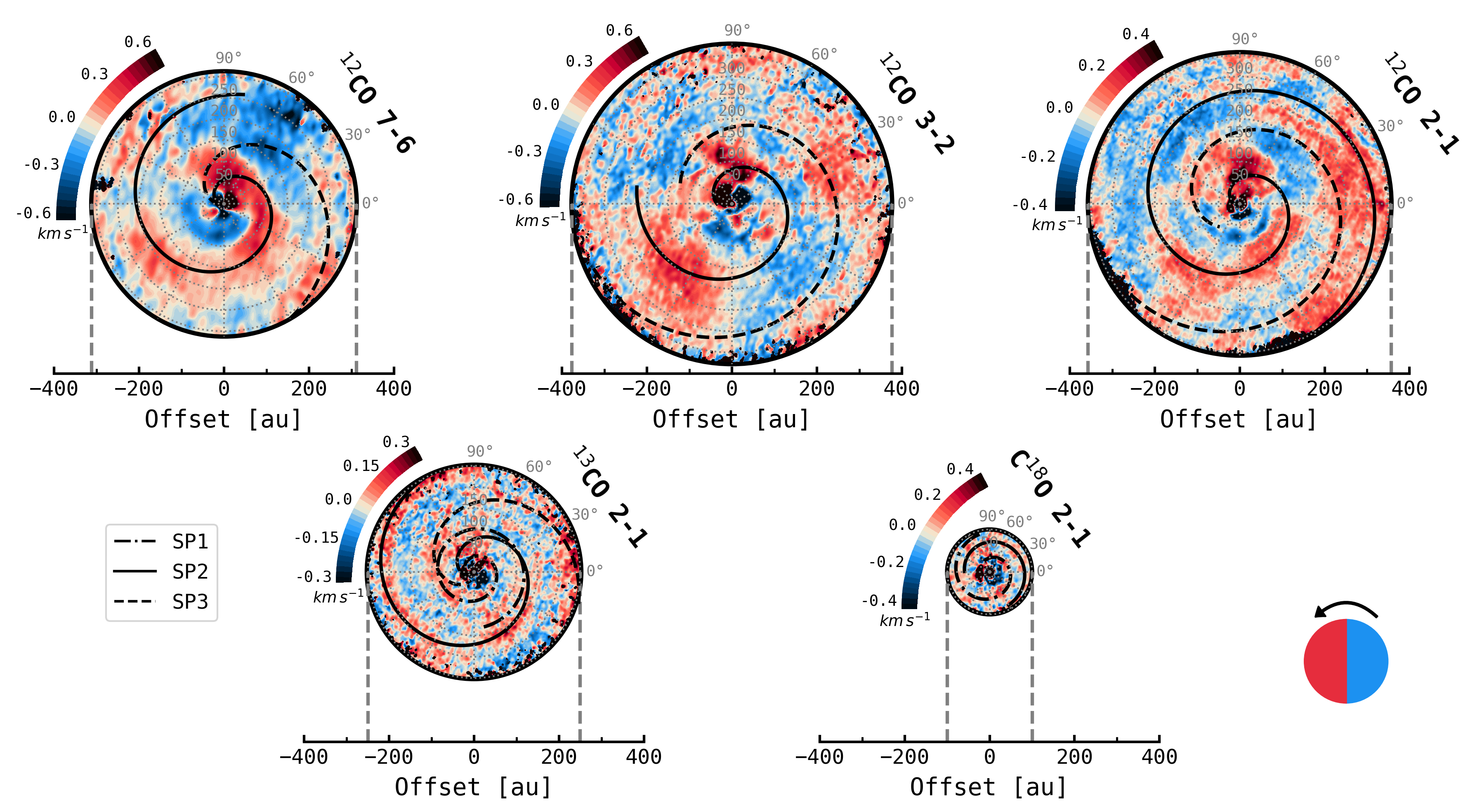}
\caption{Centroid velocity residuals, shown for the five lines studied in this work. All lines exhibit extended spiral substructures. A linear parameterization of the spirals is overlaid as black lines. The spirals are labeled depending on their (increasing) radius as SP1, SP2, and SP3. The circle in the lower right corner indicates the blue- and redshifted sides of the disk.}\label{fig:ResidualsVel}
\end{figure*}

The residuals of the different lines partly reach high values, suggesting that a smooth Keplerian model cannot fully reproduce the data. These deviations from Keplerian rotation can be used to unveil the mechanisms being at play, including massive planets but also other physical processes such as disk winds, hydrodynamical instabilities, or turbulence.
\subsubsection{Observables}
From the line intensity profiles modeled by the \textsc{discminer}, we extract three observables to search for substructures in the gas disk of HD\,100546: the line width, centroid velocity, and peak intensity or brightness temperature. These are computed from Gaussian kernels which are fitted to both the data and the model and illustrated in \hyperref[fig:Observables]{Fig.~\ref*{fig:Observables}} (left and middle panels) for the $^{12}$CO 2-1 line. The residuals resulting from the deviations of the data from the model are depicted in the right panels. The different observables can be used to trace various physical properties of the disk. The line width and peak intensity are shaped by both the gas temperature and density, the line width can additionally be used to trace turbulent motions. The gas motions which manifest in the kinematics are driven by dynamical, and often coupled, processes, including the gravitational influence of embedded (massive) bodies. Using the vertical emission structure together with the orientation of the disk as modeled by the \textsc{discminer}, the observables can be deprojected onto the disk's reference frame and their substructures can be analyzed in terms of location, magnitude, and extent. 
%
%
\section{Results}\label{sec:results}
\subsection{Kinematics}
In \hyperref[fig:ResidualsVel]{Fig.~\ref*{fig:ResidualsVel}}, we show the centroid velocity residuals for all five lines, ordered from the brightest to the faintest, and deprojected onto the disk frame. It is striking, that prominent spiral features are visible in all maps: in $^{12}$CO 7-6, a large coherent spiral runs from the inner disk around 50\,au and 90\,$\degree$ out to radii of 300\,au, and possibly further, covering more than one full azimuth. In $^{12}$CO 3-2, similar features can be discerned, yet they are not as connected. In $^{12}$CO 2-1, which has a higher spatial resolution, the spiral seems to consist of several arms which are slightly more tightly wound than in $^{12}$CO 7-6. In $^{13}$CO and C$^{18}$O, the spirals become even more tightly wound. The anchoring point of the spiral substructures lies between 50 and 150\,au and between 60 and 120\,$\degree$. 

\begin{figure*}
\centering
\includegraphics[width=1.0\textwidth]{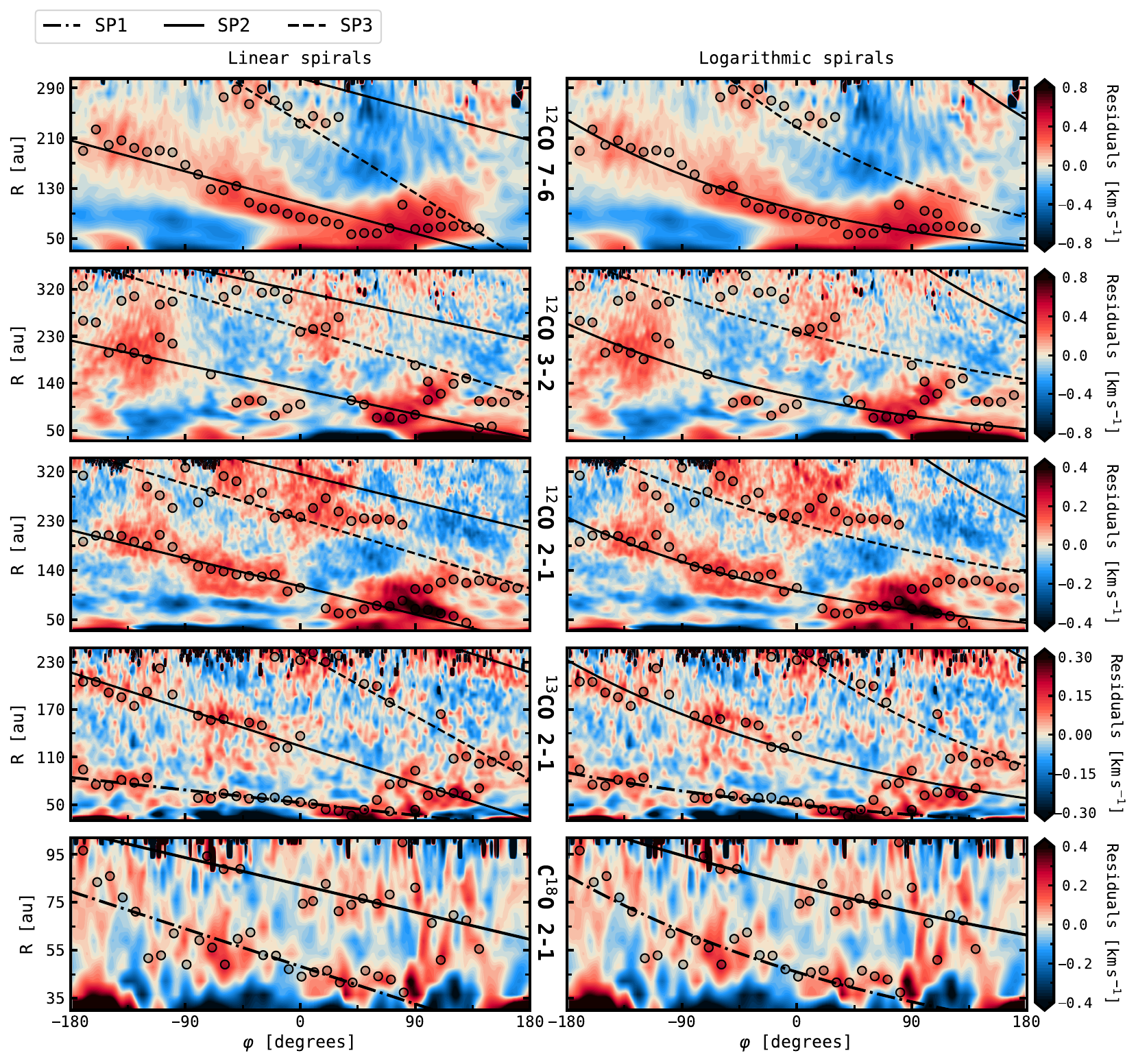}
\caption{Azimuthal deprojection of the centroid velocity residuals of the five lines, shown with overlaid linear (left) and logarithmic (right) fits of the values (circles) returned by a Gaussian fit of the  peaks in the spiral region. The spirals are labeled depending on their (increasing) radius as SP1, SP2, and SP3.}\label{fig:PolarVel}
\end{figure*}

To further analyze the spiral features, we compute the binned residuals as radius versus azimuth, presented in \hyperref[fig:PolarVel]{Fig.~\ref*{fig:PolarVel}}. For each azimuthal bin every 10\,\degree, we then fit a Gaussian to the peaks in the spirals region along the radial direction. The resulting points are overlaid in \hyperref[fig:PolarVel]{Fig.~\ref*{fig:PolarVel}}. Lastly, we fit an Archimedean or linear spiral
\begin{equation}
r = a + b\phi,  
\end{equation}
and a logarithmic spiral 
\begin{equation}
r = ae^{b\phi}, 
\end{equation}
where $r$ is the radius and $\phi$ the polar angle of the spiral, to the retrieved points. In \hyperref[fig:ResidualsVel]{Fig.~\ref*{fig:ResidualsVel}} we overlay the obtained linear spirals, the logarithmic spirals are presented in \hyperref[fig:ResidualsVelLog]{Fig.~\ref*{fig:ResidualsVelLog}}. The overall morphology of the spiral substructures is well represented by both parameterizations, however, the logarithmic spirals tend to deviate more from the features in the outer disk, while they better reproduce the bending in the inner disk than the linear spiral, which can by definition not account for that. The different spirals are labeled as SP1, SP2, and SP3 depending on their approximate radial locations (i.e., SP1 is the spiral closest in, SP3 the farthest), yet there is an offset between some of the spirals (namely SP2 in C$^{18}$O and the other lines) and it is difficult to say for sure if they are tracing the same structure. Also, some of the spirals are not visible as a connected structure. Comparing the three $^{12}$CO lines, it seems that a blue feature, separating the inner and outer parts of the spiral, is emerging more and more from $^{12}$CO 2-1 to $^{12}$CO 7-6. This suggests that other mechanisms shaping the kinematics may be coming into play, for example, interactions with the disks's environment (such as infall) which are adding a large velocity dispersion.

\begin{figure*}
\centering
\includegraphics[width=\textwidth]{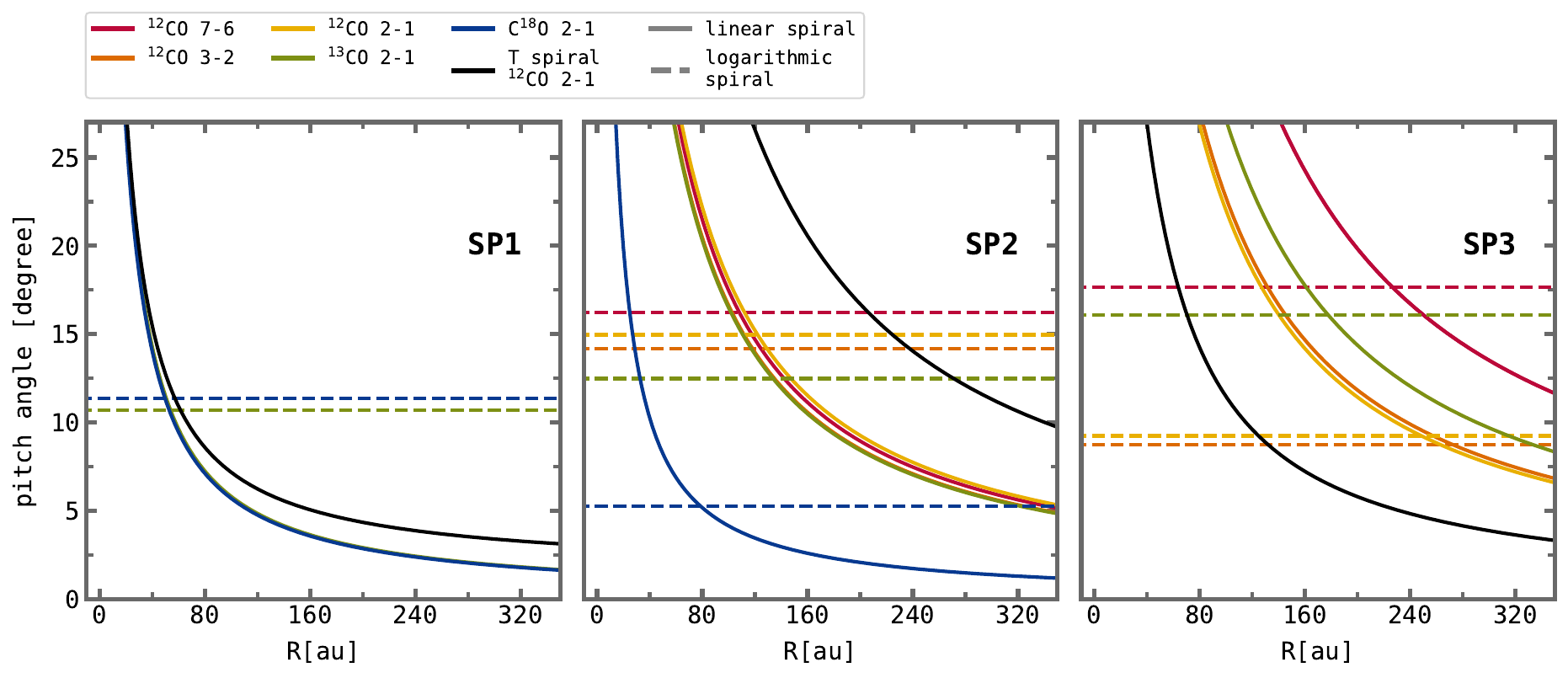}
\caption{Pitch angles of the spirals observed in the kinematics (and peak intensity of $^{12}$CO 2-1), obtained from a linear and a logarithmic fit.}\label{fig:pitch}
\end{figure*}

The parameters of the individual spiral prescriptions are presented in \autoref{tab:paramSP}. They can be used to compute the opening angle or pitch angle $\beta$ of the spirals with
\begin{equation}
\tan \beta = \Bigg|\frac{dr}{d\phi}\Bigg| \cdot \frac{1}{r}. 
\end{equation}
This results in a constant pitch angle for the logarithmic and a radially varying one for the linear spirals. The pitch angles are plotted in \hyperref[fig:pitch]{Fig.~\ref*{fig:pitch}}. For the linear spirals, the pitch angles are relatively small in the outer disk regions but rapidly increase inside of $\sim$ 50\,au. The constant pitch angles found from the logarithmic spirals have values below 20\,$\degree$. For SP1, the pitch angles are in good agreement for both $^{13}$CO and C$^{18}$O. Similarly, the linear pitch angles of SP2 align for all lines except C$^{18}$O, where we are likely tracing a different substructure. Moreover, the data quality may not be sufficient to properly analyze the spiral patterns in C$^{18}$O. The logarithmic pitch angle of SP2 in $^{13}$CO is below that of the $^{12}$CO lines. Related to that, there appears to be a trend of decreasing pitch angle with height for SP2 and SP3. This may be in agreement with the predictions of \cite{Juhasz2018} from modeling of thermal stratification, however, to properly test this scenario, the spirals would need to be traced a higher resolution and sensitivity. We further note, that the pitch angles overall decrease from SP3 to SP1. 
\begin{table*}
\centering
\small
\caption{Parameters to reproduce the spirals with linear and logarithmic functions.}\label{tab:paramSP}
{\renewcommand{\arraystretch}{1.3}%
\begin{tabular}{lllcccccc}
\toprule
\toprule
& & Parameter & $^{12}$CO 7-6 & $^{12}$CO 3-2 & $^{12}$CO 2-1 & $^{13}$CO 2-1 & C$^{18}$O 2-1 & $^{12}$CO 2-1 $T_{\mathrm{B}}$\\
\midrule
Linear & SP1 & a (au)  & - & - & - & 52.45 & 48.29 & 96.19\\
& & b ($\degree$\,au$^{-1}$) & - & - & - & --0.18 & --0.17 & --0.17\\
& SP2 & a & 107.47 & 128.18 & 110.93 & 124.01 & 82.28 & 88.17\\
& & b & --0.55 & --0.52 & --0.57 & --0.52 & --0.13 & --1.05\\
& SP3 & a & 235.79 & 246.14 & 234.18 & 242.24 & -- & 261.10\\
& & b & --1.26 & --0.73 & --0.71 & --0.89 & -- & --0.35\\
Logarithmic & SP1 & a (au)  & - & - & - & 49.95 & 45.90 & -\\
& & b ($\degree$\,au$^{-1}$) & - & - & - & --0.0033 & --0.0035 & -\\
& SP2 & a & 96.13 & 114.95 & 102.39 & 116.02 & 81.89 & -\\
& & b & --0.0051 & --0.0044 & --0.0047 & --0.0039 & --0.0016 & -\\
& SP3 & a & 227.42 & 238.05 & 226.14 & 244.09 & - & -\\
& & b & --0.0056 & --0.0027 & --0.0028 & --0.0050 & - & -\\
\bottomrule
\end{tabular}
}
\end{table*}
\begin{figure*}
\centering
\hspace{-0.8cm}
\includegraphics[width=0.38\textwidth]{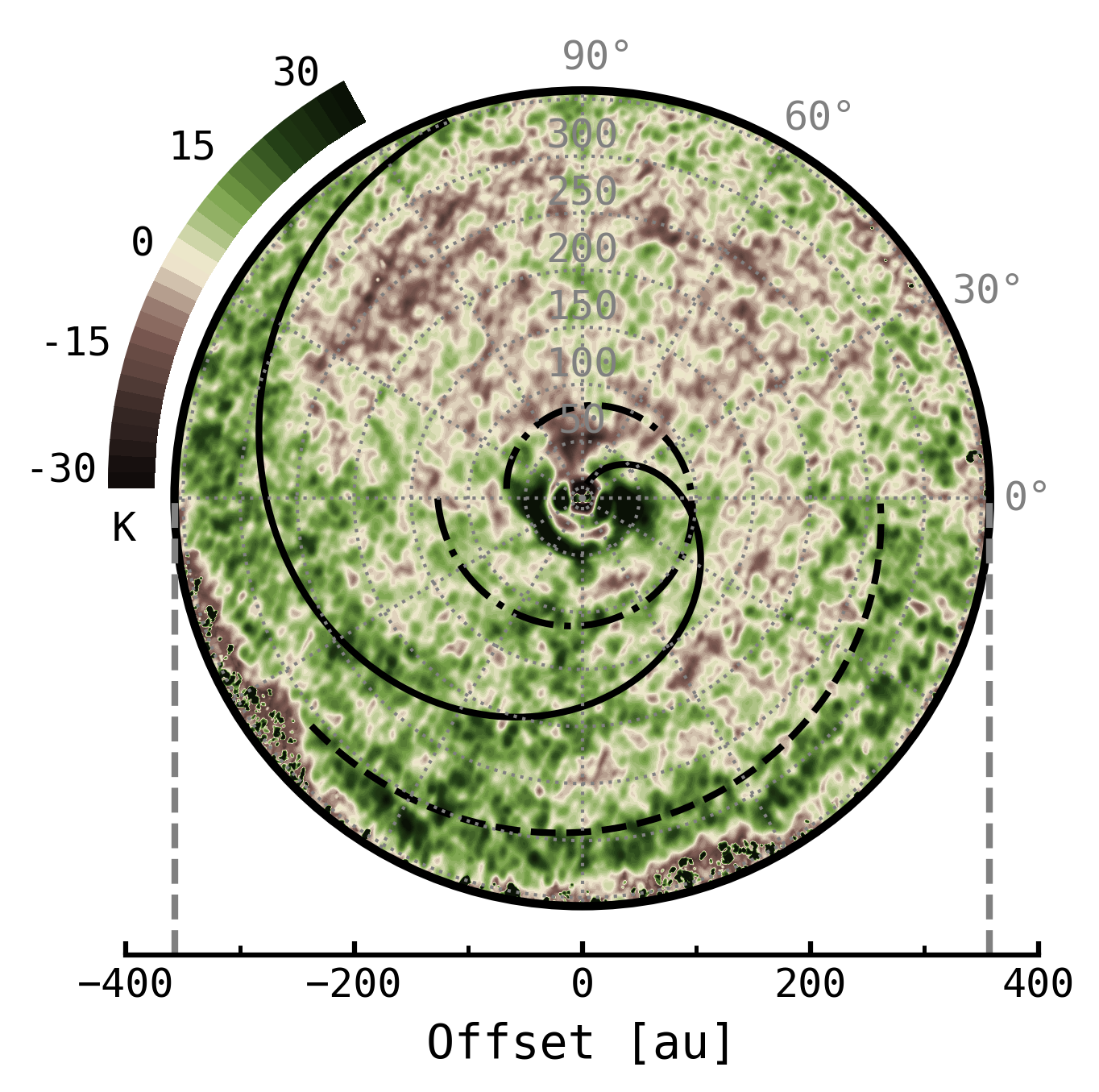} \hspace{1.cm}
\vspace*{-0.4cm}
\includegraphics[width=0.372\textwidth]{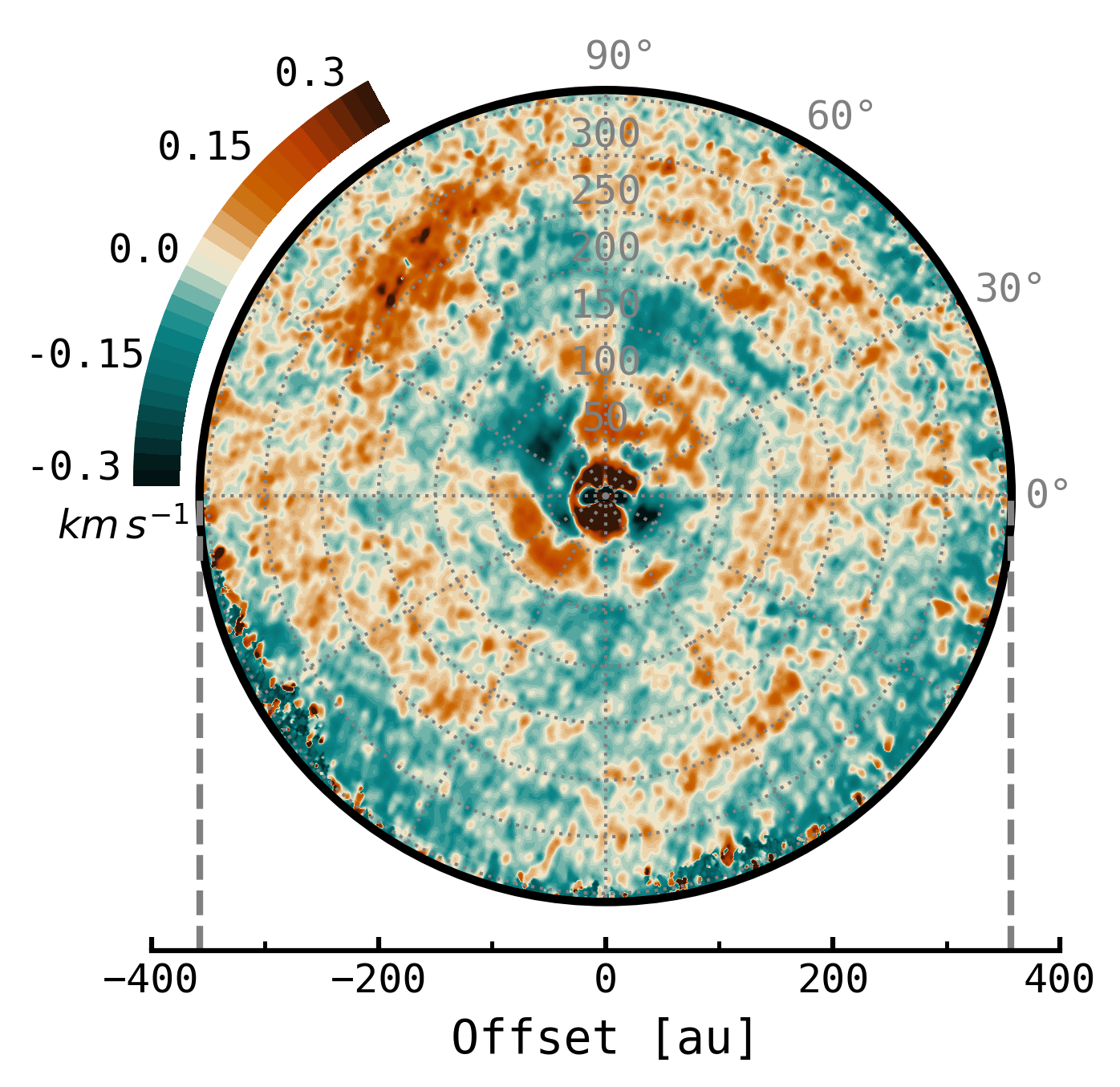}
\includegraphics[width=\textwidth]{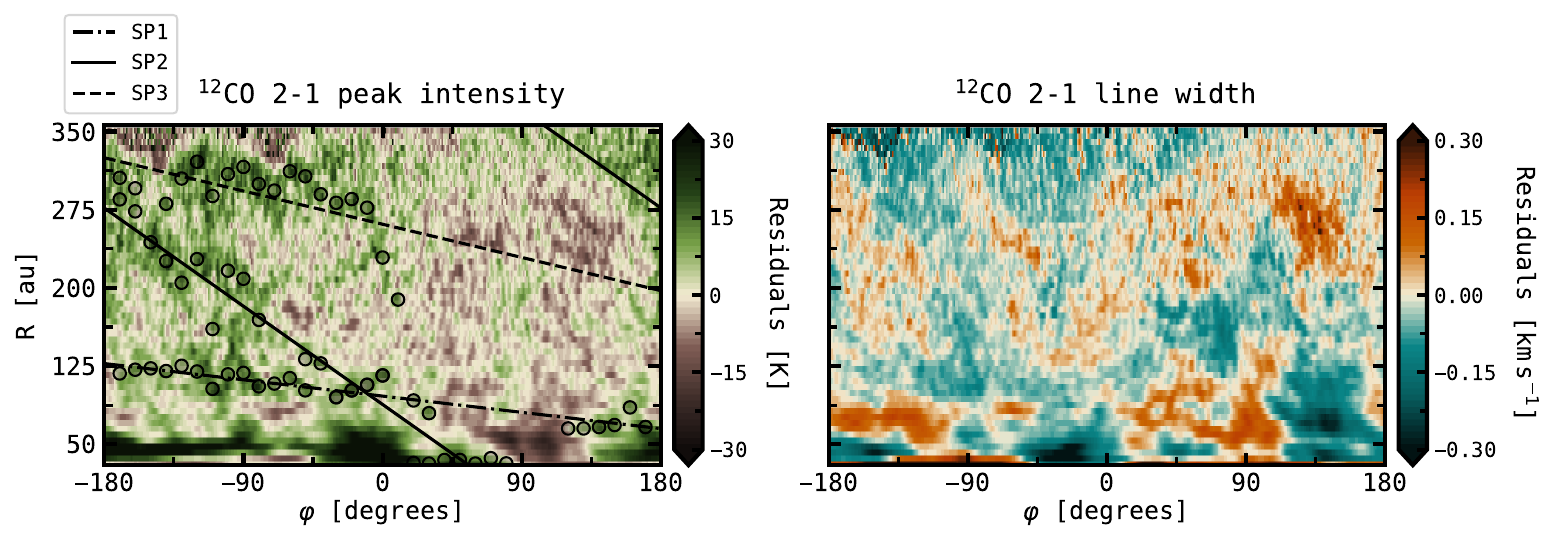}
\caption{Peak intensity (left) and line width (right) residuals of $^{12}$CO 2-1. Shown are the deprojected maps (top) and an azimuthal deprojection (bottom). In the peak intensity, three parameterizations of a linear spiral are overlaid, obtained from a Gaussian fir of the peaks in the spirals region.}\label{fig:polar12co}
\end{figure*}
\subsection{Peak intensities and line widths}
\begin{figure*}
\centering
\includegraphics[width=1.\textwidth]{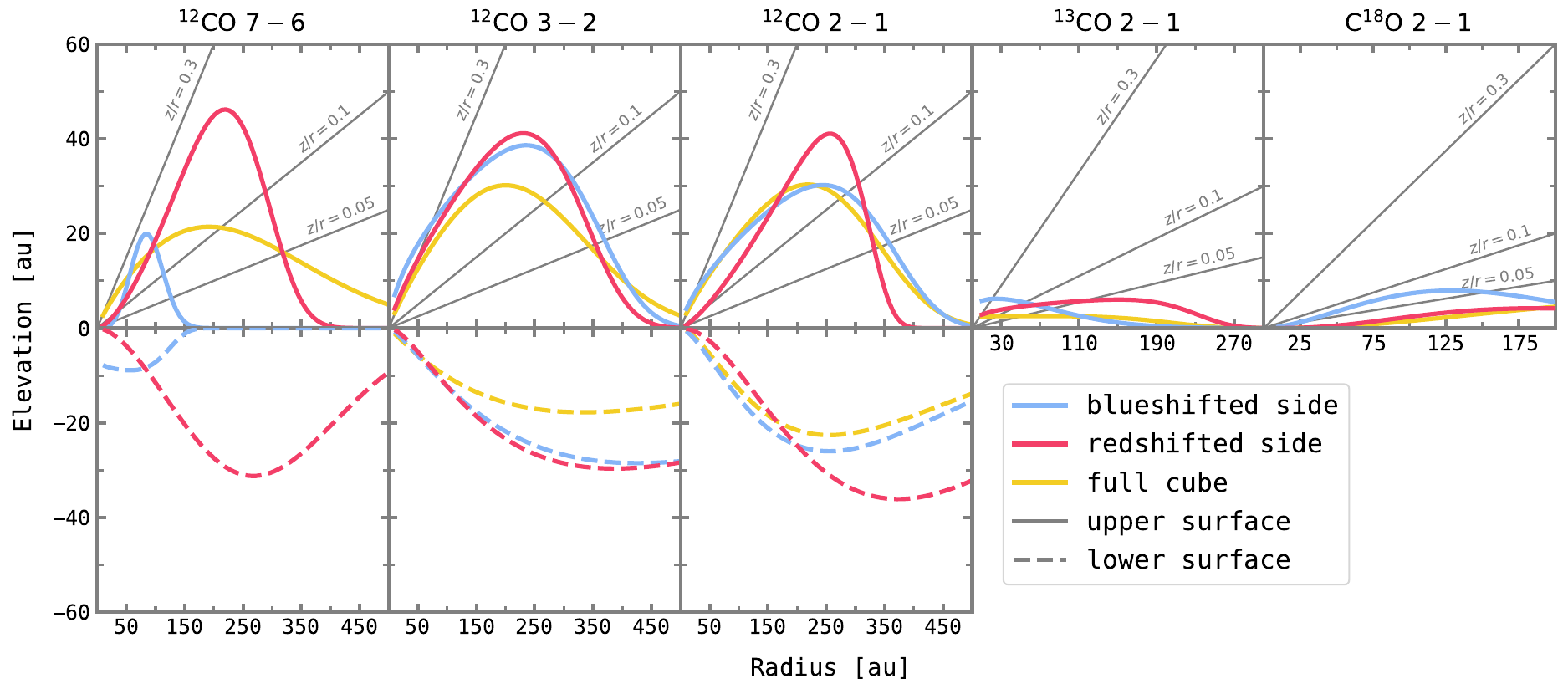}
\caption{Upper (top) and lower (bottom) emission surfaces of the five molecular lines, shown for the fits of the full cube and the fit of the blue- and redshifted sides only. }\label{fig:Heights}
\end{figure*}
The residuals of the peak intensity and line widths are displayed in the Appendix in Figs. \ref{fig:ResidualsT} and \ref{fig:ResidualsLW}, again shown ordered from the brightest to the faintest line. The substructures in the peak intensity and line width are less clear than in the kinematics and no coherent spiral structures can be discerned. In $^{12}$CO 7-6, the upper half of the disk (corresponding to the southwest side of the disk in disk coordinates) happens to be colder while the lower half is hotter. This corresponds to the brightness asymmetry mentioned before, which is seen in the channels. The line widths on the other hand are increased in the upper half of the disk, but they are decreased in the lower side. Together, this may suggest that the surface density is larger on the upper (south) side, thus broadening the line width and at the same time lowering the temperature. The additional material could be a result of infall processes, which do not only increase the density but also add a large velocity dispersion, which affects the shape of the line profile. In $^{12}$CO 3-2, a similar pattern opposite pattern (i.e.enhanced temperatures correspond to decreased line widths) is seen in the south side of the disk. Another feature is a localized temperature enhancement around 150\,au and 60\,$\degree$, which corresponds to the spot already seen in the channel maps. The features in $^{12}$CO 3-2 are very hard to interpret and better resolution data (including the short and long baseline observations with the same setup) are needed to make any quantitative remarks. Similarly, higher sensitivity data are needed for $^{13}$CO and C$^{18}$O, which show cross-like patterns in the line width residuals, resulting from the missing fit of the lower surface.

Somewhat clearer substructures can be discerned in $^{12}$CO 2-1, for which we show the projected and polar-deprojected maps in \hyperref[fig:polar12co]{Fig.~\ref*{fig:polar12co}}. In the peak intensity, three tentative spiral- or arc-like structures are visible, which we reproduce with the same methods as described above and with a linear function. The parameters and pitch angles obtained for these spirals are included in \autoref{tab:paramSP} and \hyperref[fig:pitch]{Fig.~\ref*{fig:pitch}}. Furthermore, a prominent ring of enhanced line widths can be distinguished between $\approx 125-330$\,au, showing a particularly bright region between 100--150\,$\degree$. The ring co-locates with both low and high temperatures and it is thus difficult to say if the lines are broadened by temperature effects or density enhancements. The localized region on the other hand coincides with a low-temperature spot and may be tracing a region of dense gas material or turbulent motions rather than temperature. The patterns seen in the peak intensity residuals of $^{12}$CO 7-6 and $^{12}$CO 3-2 somewhat resemble those seen in $^{12}$CO 2-1 in the south regions of the disk. Similarly the line width patters of $^{12}$CO 3-2 and $^{12}$CO 2-1 are somewhat consistent. It is possible that resolution and sensitivity effects are washing out the features in the B7 and B10 emission and that the spiral patterns seen in the peak intensity of $^{12}$CO 2-1 are simply not separated but blended together. 
\subsection{Emission heights}
In \hyperref[fig:Heights]{Fig.~\ref*{fig:Heights}}, we present the emission layers of the upper and, in case of $^{12}$CO, lower surfaces of the disk, extracted from the \textsc{discminer} models. Here we plot the profiles obtained from runs fitting for the full cube (yellow) as well as the blue- and redshifted sides only. The curves show that while the optically thick $^{12}$CO emission is coming from more elevated disk layers, particularly in the higher transitions, $^{13}$CO and $^{18}$CO originate from similar layers that are very close to the midplane. Furthermore, an asymmetry between the red- and blueshifted side is revealed, which becomes particularly strong for $^{12}$CO 7-6: the emission of the redshifted side of the disk is coming from higher disk layers ($\sim$ 25 au difference for $^{12}$CO 7-6) than that of the blueshifted side, which can already be seen in the channel maps (see for example \autoref{fig:ChannelsComp12co76both}). Such an asymmetry has been detected before only for the Elias 2-27 disk, a massive and gravitationally unstable disk, that may be affected by infall or an inner warp \citep{Paneque2021,Paneque2022a}. 

While the asymmetry between the red-and blue-shifted sides is consistently traced by the emission lines, the magnitude of the difference may vary depending on the modeling procedure. Especially for the higher J lines, the channels are twisted, which may have affected the fit of the emission surface.
\begin{figure}
\hspace{-0.35cm}\includegraphics[width=1.16\textwidth]{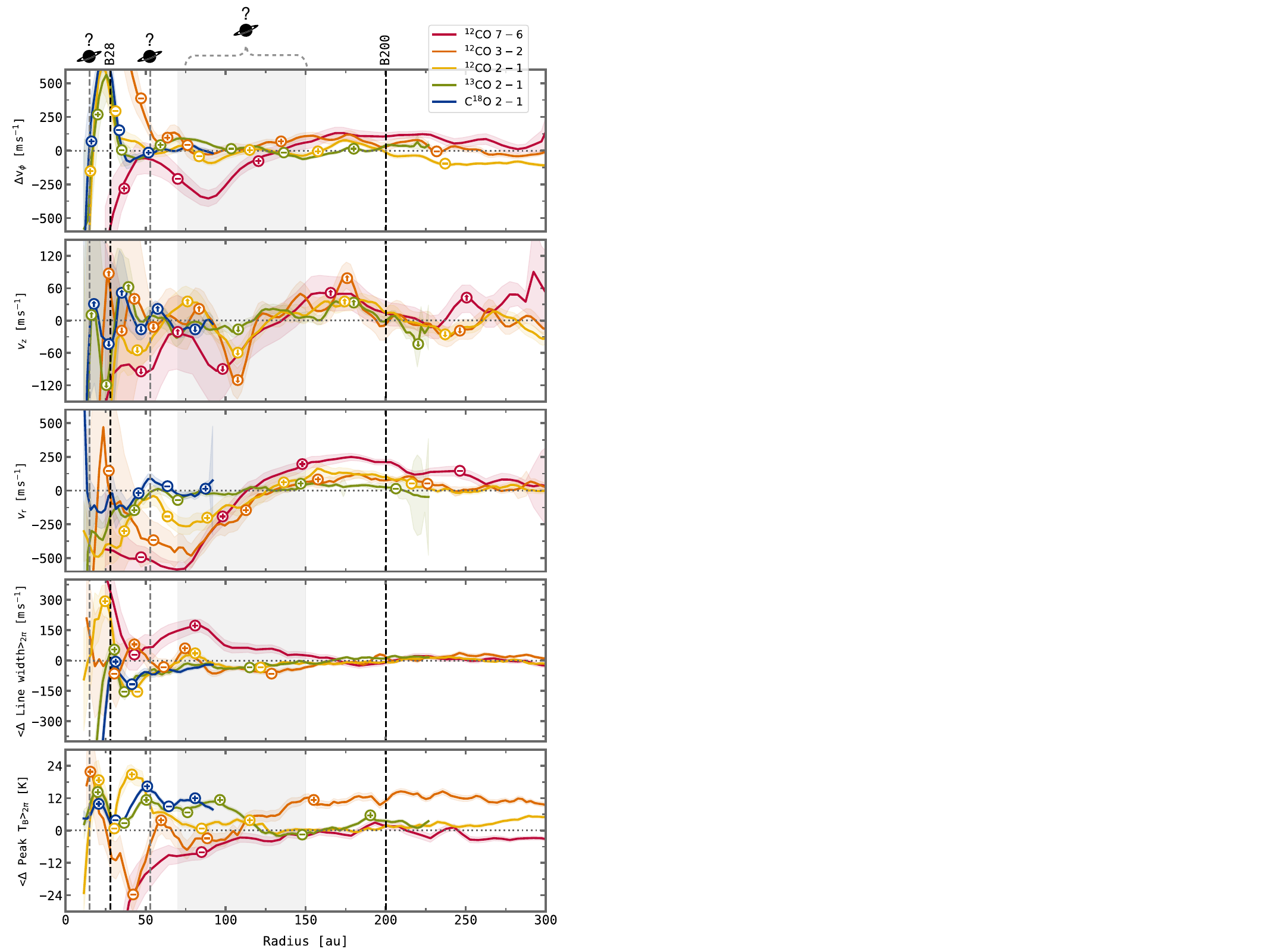}
\caption{Azimuthally averaged profiles of the three velocity components (top three panels), line width (forth panel), and peak intensity (lower panel) residuals, shown for all five lines. The centers of the dust rings and locations of previously proposed planets are marked by dashed lines and a shaded region. In the top panel, the sign of the velocity gradient is indicated with plus and minus signs. In the second row, flows hinting at gas moving away from and towards the disk midplane are highlighted as up and down arrows. The gradients of outward and inward flows are marked by plus and minus signs in the third panel}. Local line width and temperature maxima and minima are indicated with plus and minus signs in the fourth and fifth panel, respectively.\label{fig:Averages}
\end{figure}
\subsection{Azimuthal averages}
In \hyperref[fig:Averages]{Fig.~\ref*{fig:Averages}}, we plot the azimuthally averaged residuals, including all three velocity perturbations $v_{\mathrm{\varphi}}$, $v_{\mathrm{z}}$, $v_{\mathrm{r}}$, aside from the line width and peak intensity deviations. The derivation of the individual velocity components is detailed in \cite{Izquierdo2023,Izquierdo2025}. In \hyperref[fig:Averages]{Fig.~\ref*{fig:Averages}}, we mark the locations where planets have previously been proposed as well as the centers of the two observed dust rings (B28 and B200). We further annotate some trends seen in the profiles such as the sign of the velocity gradient in the azimuthal perturbations (tracing pressure minima and maxima), upwards and downwards flows in the vertical perturbations, negative and positive gradients in the radial velocity, and local minima or maxima in the line width and peak intensity. We discuss some of the trends seen in the averages in the following, however, we note that their computation is based on the assumption of axisymmetry, which does not necessarily hold. In fact, the spiral structures seen in the kinematics suggest the perturbations to be non-symmetric. Nevertheless, the curves are still useful to get an indication of some general trends.   

At the location of the dust rings, negative velocity gradients are seen in all lines except $^{12}$CO 7-6, tracing pressure bumps which can account for the formation of the rings. In this context, the more optically thin lines tracing regions closer to the midplane can better constrain the underlying density profile than the optically thick lines which are affected by the temperature. Thus, it is not surprising that $^{12}$CO 7-6 is showing an opposite trend. Inside the inner dust cavity, a strong positive gradient is present and downward flows indicate the gas material falling into the cavity. However, close to the star, beam smearing and optical depth effects are becoming significant, resulting in large errors, and inside of $\sim$ 25\,au the profiles have to be treated with caution. For the same reason, we are not able to draw any conclusions on the features seen around the innermost planet candidate at $\sim$ 15\,au. Around the location of the second planet candidate at $\sim$ 53\,au, both $^{13}$CO and C$^{18}$O show positive gradients, supporting the presence of a gas gap that indeed may be carved by a massive companion. Again, an opposite trend is visible in $^{12}$CO but as mentioned above, $^{13}$CO and C$^{18}$O are more likely to trace pressure minima related to a low gas density. Just inside the planet candidate's location, downward flows are seen in all lines, further supporting the presence of a companion. Moreover, there are line width minima that align well with those downward flows. Between 50 and 150\,au, a change of the gradient of the radial velocity is observed in all lines, corresponding to inward motions inside of the $\sim 100$\,au and outward motions outside of $\sim 100$\,au. The brightness temperature shows strong fluctuations inside of $\sim$ 100\,au, which are hard to put into context with the other substructures and their origin cannot really be assessed.

Except for C$^{18}$O, for which the emission is not as extended, positive azimuthal velocity gradients and significant downward flows are present in all molecular tracers between 75--125\,au, which coincides with the dust gap found by \cite{Fedele2021}, the third proposed location of a planet candidate, the location of the bright spot in $^{12}$CO 3-2, and the gap seen in the radial profile of $^{13}$CO. For most of the radii, where the ring of enhanced line widths is observed in $^{12}$CO 2-1 (see \autoref{fig:polar12co}), a negative azimuthal velocity gradient is found, supporting that we are indeed tracing a gas ring.   
\section{Discussion}\label{sec:discussion}
\subsection{On the planet candidate at 50\,au}
All molecular tracers analyzed in this study exhibit extended spiral structures in the kinematical residuals, which we reproduce with linear and logarithmic functions (\hyperref[fig:ResidualsVel]{Figs.~\ref*{fig:ResidualsVel}}, \ref{fig:PolarVel}). Their consistency suggests them to be real features tracing a common origin. The pitch angles of the linear spirals show small values at most radii but rapidly increase inside of $\sim$50\,au. Such a behavior is consistent with Lindblad-resonance-driven spiral wakes of a massive embedded companion (e.g., \citealp{Ogilvie2002,Rafikov2002,Bae2018a,Bae2018b}), which tend to increase only close to the location of the planet. The logarithmic spirals on the other hand show constant pitch angles below $20\,\degree$, which are more consistent with tightly wound spirals excited by a planet through buoyancy resonances \citep{Bae2021}. In both cases, the spirals support the presence of a companion inside of $\sim$ 50\,au and therefore provide additional evidence for the planet candidate HD\,100546\,b, even though also the candidate HD\,100546\,c could in principle explain the formation of the spirals. Furthermore, the possibility of a binary rather than a planetary companion, as invoked by \cite{Norfolk2022}, cannot be ruled out with the present data. The HD\,100546 disk has previously been suggested to be warped in the center \citep{Walsh2017}, and a binary companion would provide a good explanation for both the extended spiral structures and the misaligned inner disk. Hydrodynamical simulations are needed to further distinguish between the Lindblad and buoyancy scenario and to make predictions on the companion mass and location.

While velocity spirals do not necessarily align with spiral features in the peak intensity, it is still puzzling that except for $^{12}$CO 2-1, no counterpart of the kinematical spirals is found. Both in TW\,Hydra \citep{Teague2019Spiral,Teague2022,Teague2022a} and CQ\,Tau \citep{Woelfer2021}, temperature spirals have been observed that overlap at least partly with the spirals in the velocity residuals, suggesting that there may be a physical mechanism behind their connection. In the companion scenario, the spiral density waves lead to an increase in surface density and thus to a higher CO opacity. The $\tau =1$ layer is moved to a larger altitude, where the temperature is generally higher, resulting in the observed spiral substructure in the gas temperature \citep{Phuong2020b,Phuong2020a}. The lack of temperature features in HD\,100546 is therefore unexpected, however, velocity spirals are more easily detected than those in the peak brightness temperature. The absence could therefore result from sensitivity effects and needs to be further investigated in the future.   

Aside from dynamical interactions with a companion, spiral structures can also be a signpost of gravitational instability (GI; e.g., \citealp{Rice2003}). In that case, the pitch angles are expected to be comparable for the surface and the midplane layers, which are heated by shocks. On the other hand, the pitch angle is expected to increase towards the surface layers in the companion scenario if the disk is passively heated with a positive vertical temperature gradient \citep{Juhasz2018}. There is a tentative trend, that the spirals in HD\,100546 become more tightly wound (lower pitch angle) towards the midplane, which would be supporting the predictions by \cite{Juhasz2018} and dynamical interactions rather than GI as the underlying mechanism. However, at the current data quality it is difficult to properly test this scenario. Higher resolution and sensitivity would be needed to better resolve the spirals and trace them radially further for $^{13}$CO and C$^{18}$O. 
\subsection{On the planet candidate between 70--150\,au}
To explain the ringed substructures in the dust, a planet orbiting at large separations between 70--150\,au has previously been proposed \citep{Walsh2014,Pinilla2015,Fedele2021,Pyerin2021}. In this region, we observe vertical motions directed down towards the midplane around $\sim$ 110\,au in $^{12}$CO 3-2, $^{12}$CO 2-1, and $^{13}$CO 2-1, and around $\sim$ 95\,au in $^{12}$CO 7-6 (\hyperref[fig:Averages]{Fig.~\ref*{fig:Averages}}, panel two). This shift is not surprising as the lines trace different heights in the disk which are governed by different physical and chemical processes. Such vertical flows are commonly associated with meridional flows around forming planets and similar motions have been observed before by \cite{Teague2019a}, possibly tracing a depletion of gas material carved by an embedded companion. In that case, the gas is generally expected to radially flow towards the planet location, however, depending on the traced scale height, either inward or outward flows may be expected probing different parts of the meridional circulation. It appears that we are tracing the opposite behavior in $v_r$ around the downward vertical motions, yet the expected trend is seen further in at smaller radii. Moreover, while the vertical motions are expected to behave coherently across different scale heights as we see here, this is different for radial motions, which makes it difficult to put the $v_r$ pattern into context with a meridional flow \citep{Teague2019a}. Additionally, our estimates of the radial velocity flow may be biased by the complex spiral patterns in the disk.

The presence of a gas gap around 90--150\,au is further supported by the positive sign of the azimuthal velocity gradient at these radii (\hyperref[fig:Averages]{Fig.~\ref*{fig:Averages}}), and tentative dips of emission in the azimuthally averaged radial intensity profiles of $^{13}$CO ($\sim$ 150\,au) and C$^{18}$O ($\sim$ 85\,au) (\hyperref[fig:RadialProfiles]{Fig.~\ref*{fig:RadialProfiles}}). Lastly, a bright, localized spot stands out in the velocity channels and peak intensity map of $^{12}$CO 3-2 around 125 and 150\,au, respectively. Altogether, these features strongly support the presence of a massive planet in the outer disk. However, higher sensitivity data at high resolutions are needed to confirm the robustness of the substructures. Moreover, a comparison with hydrodynamical models is crucial to understand if they can indeed be explained by a massive embedded body.     
\subsection{Origin of the asymmetry in emission heights}
The CO emission of HD\,100546 is marked by an asymmetry between the blue- and redshifted halves of the disk, which is especially pronounced in $^{12}$CO 7-6 (\hyperref[fig:Heights]{Fig.~\ref*{fig:Heights}}). Such an asymmetry has previously been detected only for the Elias 2-27 disk \citep{Paneque2021,Paneque2022a}. One explanation could be given by infall onto the redshifted side of the disk. Infalling material results in an increase of the surface density, and thus the emission becomes optically thick in higher disk layers. Moreover, the temperature is altered as shocks will heat the layers on which the material is accreted \citep{Hennebelle2017}. Observations of scattered light have indeed revealed that the disk is surrounded by a diffuse envelope \citep{Grady2001,Ardila2007}. Another scenario that may account for the asymmetry is the presence of a warped inner disk casting a shadow over part of the outer disk, shielding it from stellar radiation and thus resulting in an azimuthal variation of the temperature, which has an immediate effect on the vertical structure of the disk. However, in the disk surface, which has a low surface density, the temperature decrements caused by a misaligned disk are expected to be smoothed out by thermal disk radiation \citep{Casassus2019a}. Evidence for a misaligned inner disk is presented in \cite{Walsh2017}, and the analysis by \cite{Bohn2022}, who compare the position angle and inclination of the inner and the outer disk, places HD\,100546 among the less aligned objects, yet, the scattered light images do not show signs of significant shadowing from a misaligned disk \citep{Garufi2016}. A warped inner disk can in principle be triggered by binary interactions or a misaligned planet (e.g., \citealp{Nealon2018}) but also the infall of material may produce such a structure \citep{Bate2010,Sakai2019}. We note again, that even though the asymmetry is consistent across the emission lines, the magnitude of the difference may be enhanced due to the modeling procedure. Connected to that, it is surprising that there appears to be no strong asymmetry between the two sides in the scattered light images. The asymmetry should therefore be further investigated with higher sensitivity data.
\section{Summary}\label{sec:summary}
In this work, we have modeled the molecular line emission of five CO lines observed with ALMA in the circumstellar disk around HD\,100546. Our main results are summarized as follows.
\begin{itemize}
\item Extended spiral features are resolved in the kinematics of all five lines. The peak intensity also shows indications of spirals but only in $^{12}$CO 2-1, likely due to limited sensitivity for the other lines. A ring and bright region of enhanced line widths are seen in the line width residuals of $^{12}$CO 2-1.
\item The spirals are well reproduced by a linear and/or a logarithmic function and show small pitch angles that are consistent with Lindblad or Buoyancy spirals driven by a companion.
\item A rapid increase of the pitch angles inside of $\sim$ 50\,au suggests an upper limit for the companion location around this radius, which is consistent with the planet candidate HD\,100546 b or HD\,100546 c but may point to a binary companion rather than a planet as suggested by \cite{Norfolk2022}.   
\item
Several indications for a companion at larger separations (90--150\,au) are present: downward flows towards the midplane and pressure minima around this region suggest a gas-depleted ring that may be carved by an unseen planet. In the radial curves of the integrated intensity of $^{13}$CO 2-1 and C$^{18}$O 2-1, tentative gas gaps are visible at 150 and 85\,au, respectively, but they need to be confirmed with higher sensitivity data. Altogether, these features coincide with the dust gap observed by \cite{Fedele2021}.  
\item The emission from the redshifted side appears more elevated compared to the blueshifted side of the disk. Such an asymmetry may result from infalling material on one side of the disk, thus increasing the density and lifting the optically thick emission layer. Temperature effects resulting from a warped inner disk casting a shadow over the disk represent an alternative explanation. 
\end{itemize}
High-sensitivity multi-line and multi-isotopologue data (of different molecules) taken at high spectral and spatial resolution are of paramount importance to probe the full vertical and radial extent of protoplanetary disks. Mapping out their structures and kinematics may enable us to distinguish different disk-shaping mechanisms, including planet--disk interactions, and to link their properties to those of the mature exoplanet populations.   
\begin{acknowledgements}
We thank the referee Simon Casassus for the constructive feedback, which greatly improved the quality of this paper. This paper makes use of the following ALMA data: 2011.0.00863.S, 2015.1.00806.S, 2016.1.00344.S, and 2018.1.00141.S. ALMA is a partnership of ESO (representing its member states), NSF (USA) and NINS (Japan), together with NRC (Canada) and NSC and ASIAA (Taiwan) and KASI (Republic of Korea), in cooperation with the Republic of Chile. The Joint ALMA Observatory is operated by ESO, auI/NRAO and NAOJ.
\end{acknowledgements}
%
%
\bibliographystyle{aa}
\bibliography{bibliography}
\begin{appendix}
\onecolumn
\section{Comparison of the $^{12}$CO 3-2 data sets}\label{appendix:chans12co32}
Here, we compare the individual short and long-baseline datasets of $^{12}$CO 3-2 to the combined data to assess if artificial features may have been introduced in the imaging process. The overall patterns are consistent among the dataset, supporting the robustness of our results.
\begin{figure}[h!]
\centering
\includegraphics[width=0.8\textwidth]{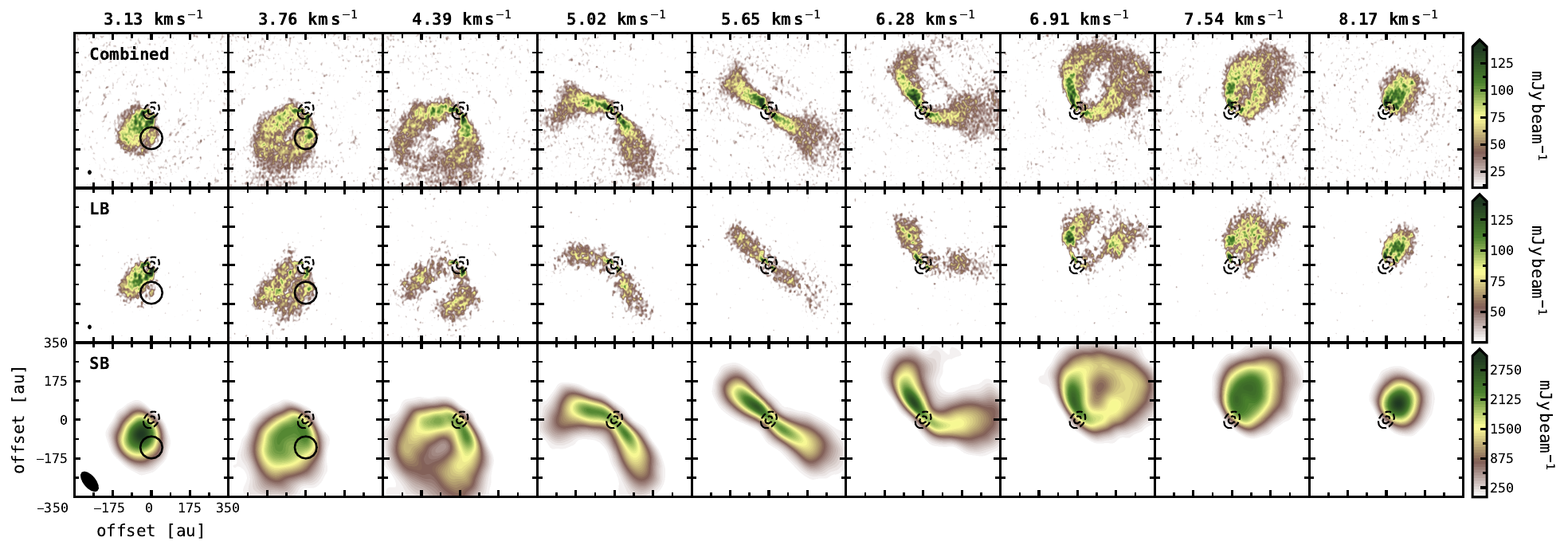}
\caption{Comparison of the different $^{12}$CO 3-2 data sets, with the combined data in the top, the long-baseline data in the middle, and the short-baseline data in the bottom row. Shown are the channels in steps or three from the central channel and for a velocity resolution of 210\,m\,s$^{-1}$. The inner millimeter continuum ring is overlaid as dashed contours. A localized feature is seen in some of the channels within the solid circle. The observational beam is shown in the bottom left corner of the first panel of each row.}\label{fig:Channels12co32}
\end{figure}
\begin{figure*}[h!]
\centering
\includegraphics[width=0.7\textwidth]{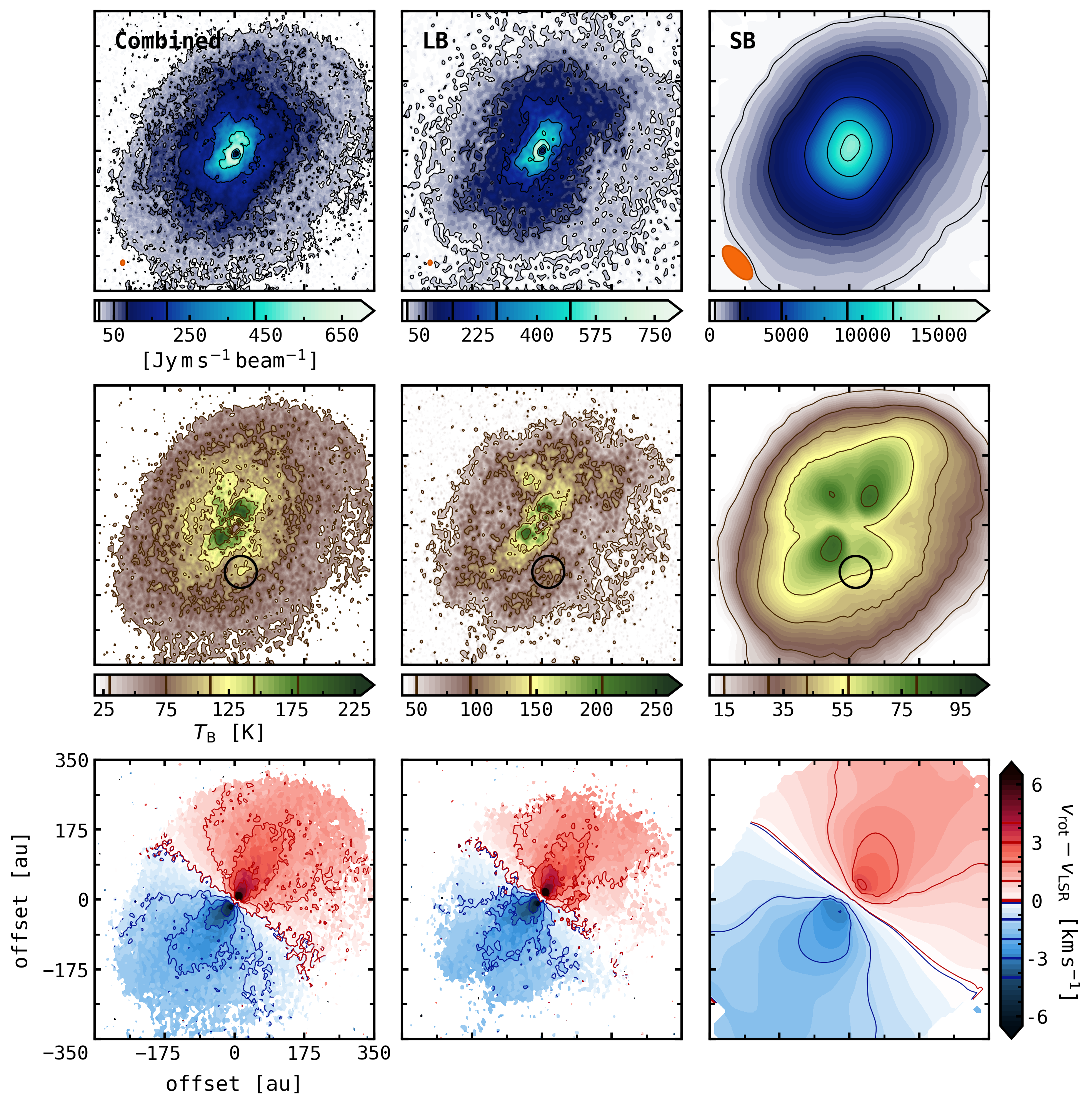}
\caption{Comparison of the different $^{12}$CO 3-2 data sets, with the combined data in the left, the long-baseline data in the middle, and the short-baseline data in the right column. Shown are the integrated intensity (top), peak intensity (middle), and line-of-sight velocity (bottom). A localized feature is highlighted by the solid circle. Some contours are overlaid and their levels are indicated in the color bars. The observational beam is shown in the bottom left corner of the first-row panels. }\label{fig:Moments12co32}
\end{figure*}
\clearpage
\section{Full data cubes}\label{appendix:chan_all}
In this section, we report the full data cubes used in the analysis, showing the whole intensity range.
\begin{figure*}[!h]
\centering
\includegraphics[width=1.0\textwidth]{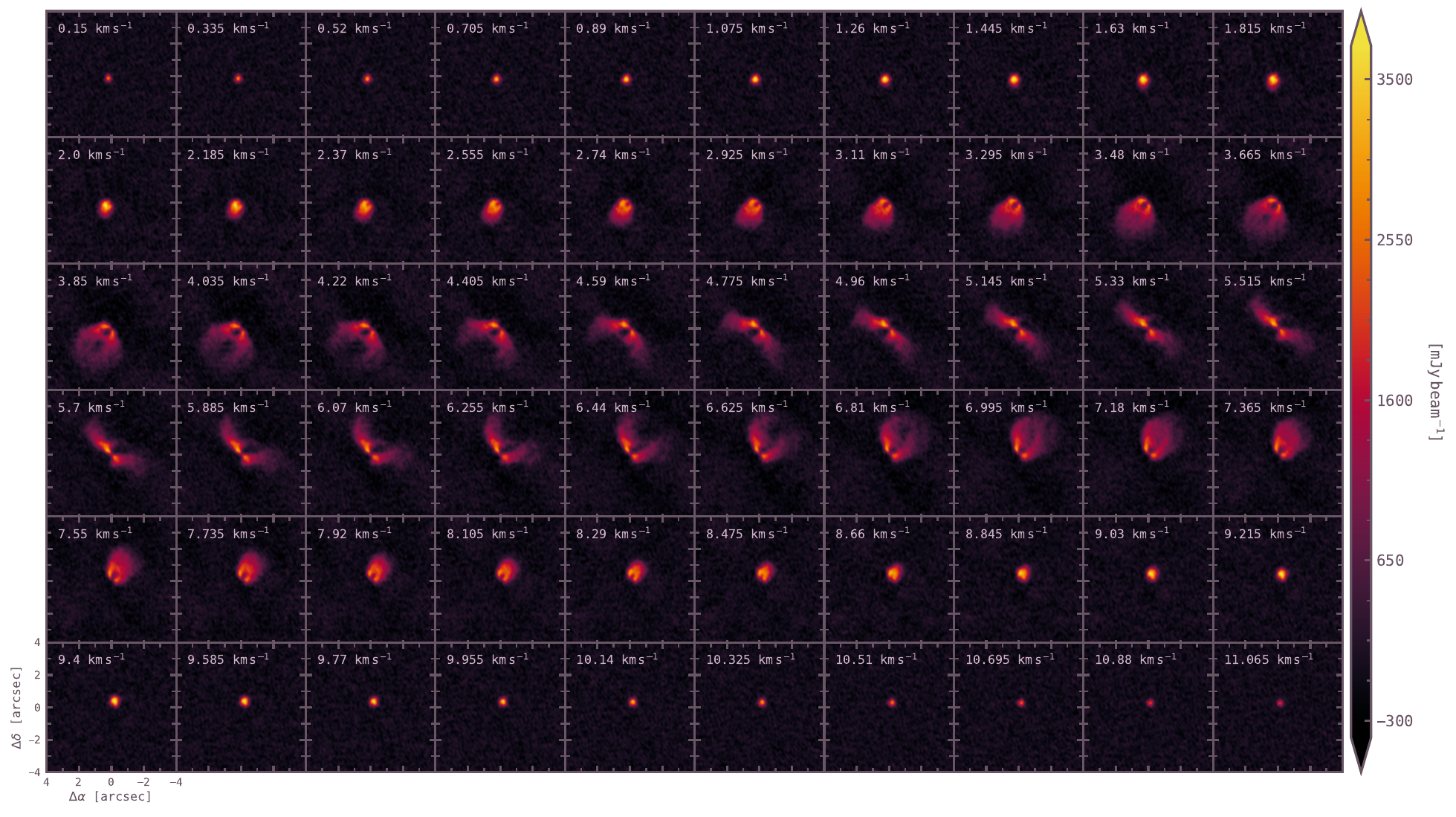}
\caption{Intensity maps of the full data cubes, shown for the $^{12}$CO 7-6 emission line.}\label{fig:Channels12co76all}
\end{figure*}
\begin{figure*}[!h]
\centering
\includegraphics[width=1.0\textwidth]{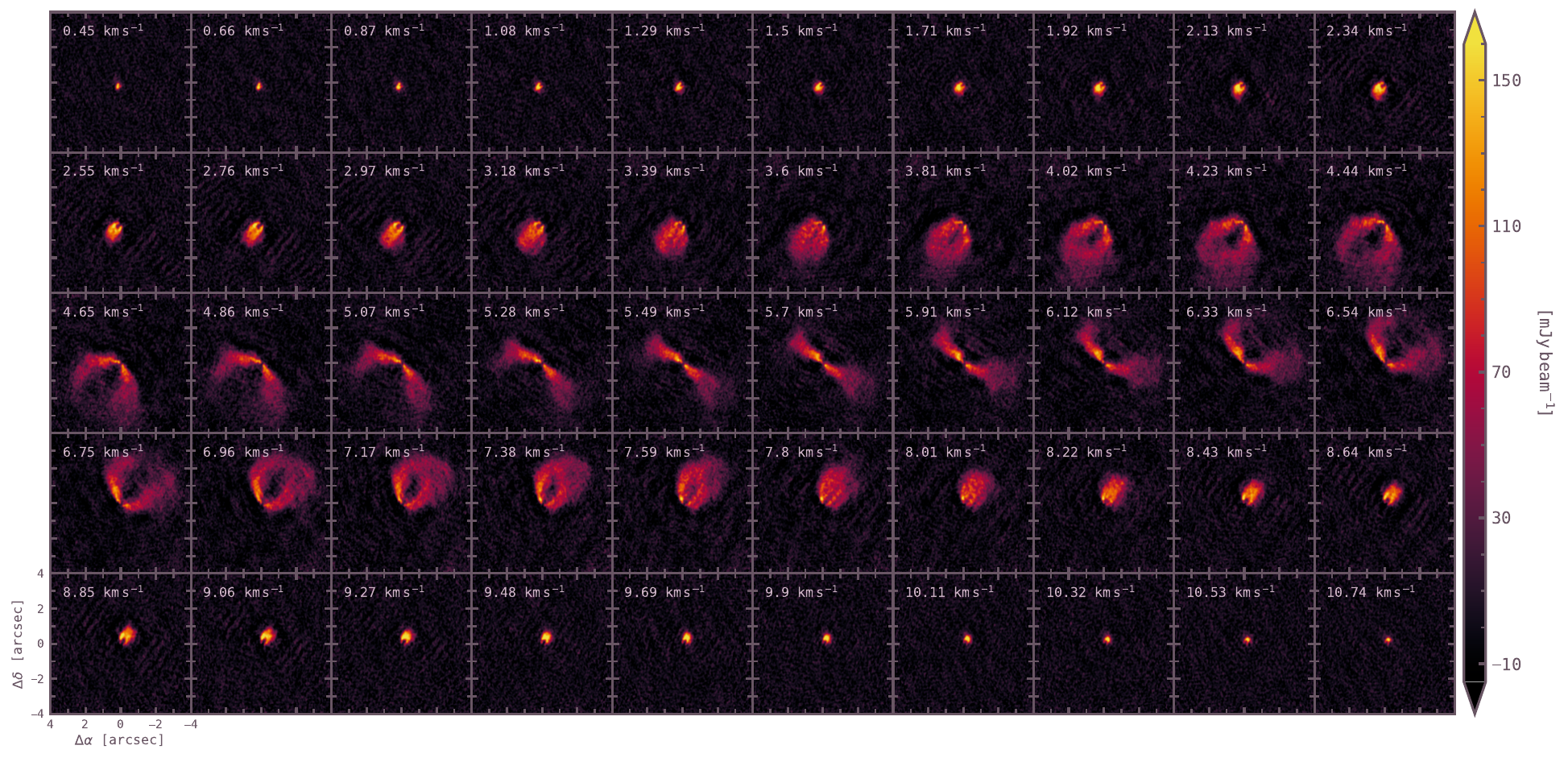}
\caption{Intensity maps of the full data cubes, shown for the $^{12}$CO 3-2 emission line.}\label{fig:Channels12co32all}
\end{figure*}
\clearpage
\begin{figure*}[!h]
\centering
\includegraphics[width=1.0\textwidth]{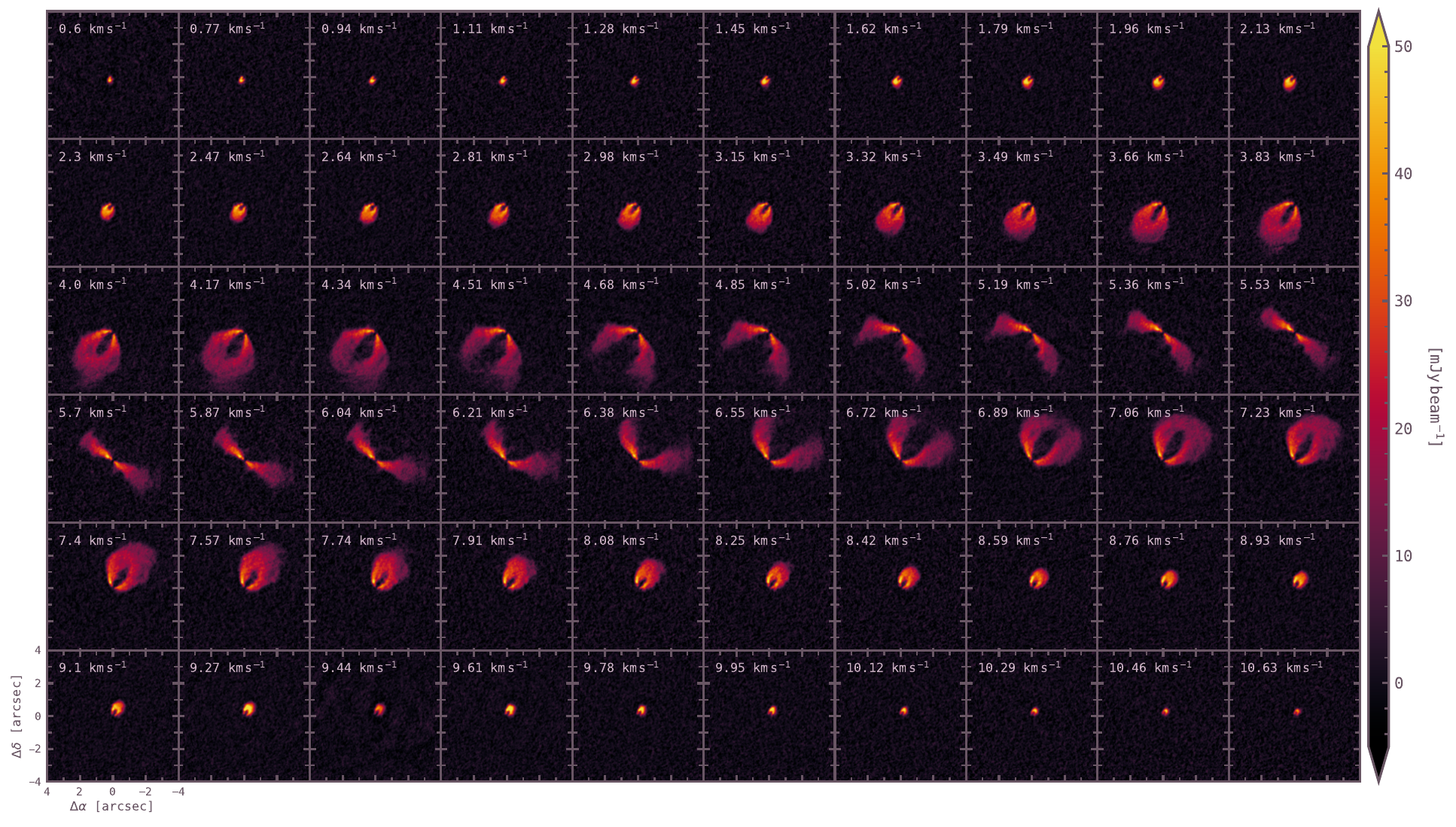}
\caption{Intensity maps of the full data cubes, shown for the $^{12}$CO 2-1 emission line.}\label{fig:Channels12co21all}
\end{figure*}
\begin{figure*}[!h]
\centering
\includegraphics[width=1.0\textwidth]{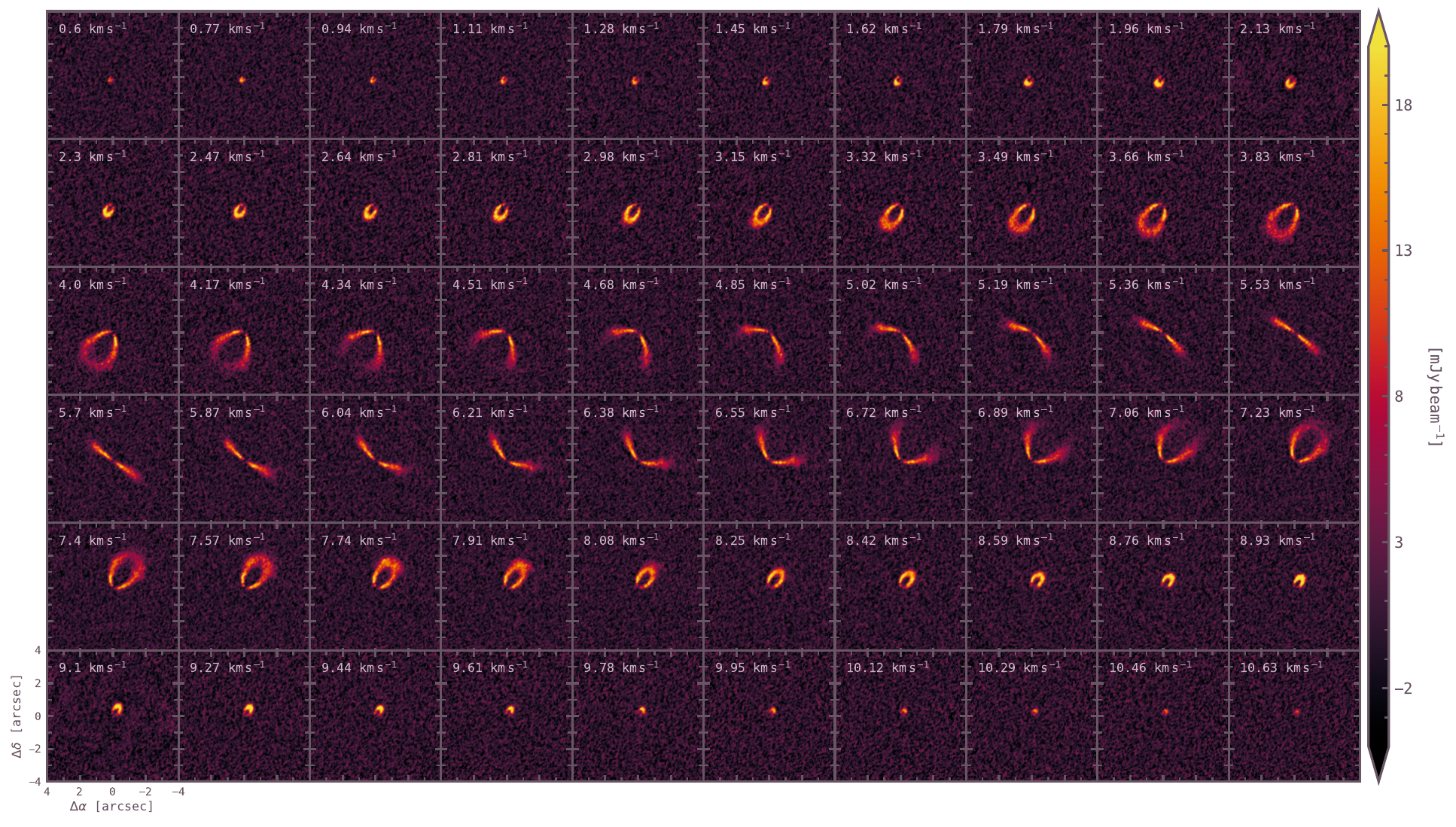}
\caption{Intensity maps of the full data cubes, shown for the $^{13}$CO 2-1 emission line.}\label{fig:Channels13co21all}
\end{figure*}
\clearpage
\begin{figure*}[!h]
\centering
\includegraphics[width=1.0\textwidth]{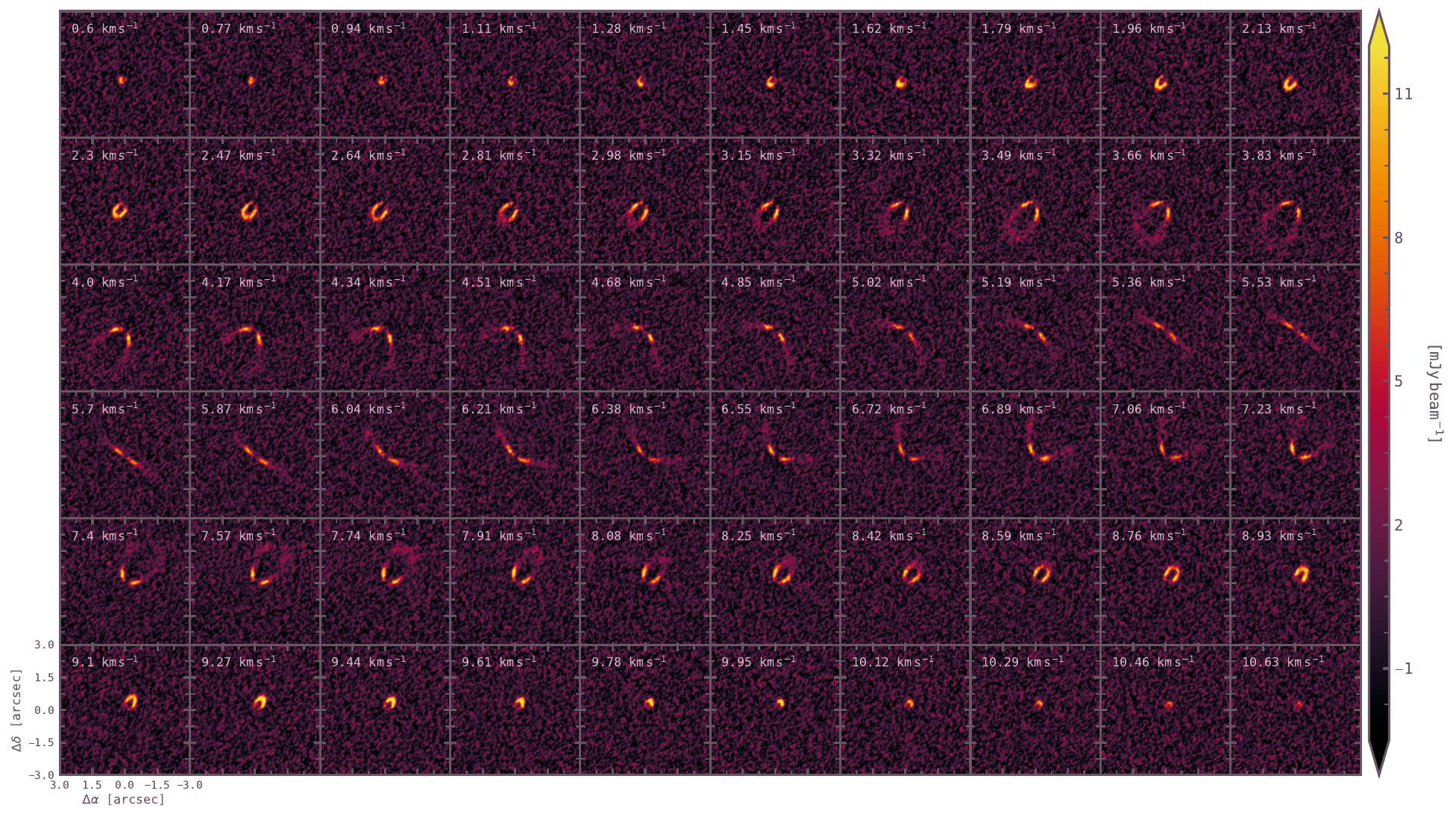}
\caption{Intensity maps of the full data cubes, shown for the C$^{18}$O 2-1 emission line.}\label{fig:Channelsc18o21all}
\end{figure*}
%
%
\section{Best-fit parameters for the blue and red-shifted cube}
\begin{table}[h!]
\small
\centering
\caption{Best-fit results of the modelling of the blueshifted channels with the \textsc{discminer}\tablefootmark{(a)}.}\label{tab:fitsblue}
\begin{tabular}{lccccccc}
\toprule
\toprule
Attribute &  Parameter & Unit & $^{12}$CO 7-6 & $^{12}$CO 3-2 & $^{12}$CO 2-1 & $^{13}$CO 2-1 & C$^{18}$O 2-1 \\
\midrule
\multirow{2}{*}{Orientation} & $x_{\mathrm{c}}$ & au & 1.03 & 7.58 & 0.25 & 3.53 & 1.97 \\
& $y_{\mathrm{c}}$ & au & 8.16 & --10.59 & --2.52 & --0.85 & 1.33 \\
\midrule
\multirow{2}{*}{Velocity} & $M_{*}$ & M$_{\odot}$ & 2.13 & 2.01 & 2.06 & 2.15 & 2.40 \\
& v$_{\mathrm{sys}}$ & km\,s$^{-1}$ & 5.45 & 5.66 & 5.69 & 5.66 & 5.441\\
\midrule
\multirow{4}{*}{Upper surface} & $z_{0}$ & au & 149.17 & 26.44 & 19.00 & 8.43 & 13.41 \\
& $p$ & - & 3.60 & 0.58 & 0.67 & 0.16 & 1.66 \\
& $q$ & - & 2.58 & 4.99 & 4.99 & 1.74 & 1.76 \\
& $R_{\mathrm{t}}$ & au & 74.08 & 362.308 & 364.39 & 99.84 & 132.99 \\
\midrule
\multirow{4}{*}{Lower surface} & $z_{0}$ & au & 9.49 & 21.53 & 24.09 & 21.12 & - \\
& $p$ & - & 0.08 & 1.42 & 1.57 & 1.41 & - \\
& $q$ & - & 4.985 & 0.80 & 1.13 & 1.89 & - \\
& $R_{\mathrm{t}}$ & au & 129.05 & 206.59 & 189.80 & 1.89 & - \\
\midrule
\multirow{6}{*}{Intensity} & $I_{0}$ & Jy\,px$^{-1}$ & 1.57 & 0.99 & 0.99 & 0.05 & 0.005 \\
& $p_0$ & - & 0.94 & 0.32 & --0.45 & 1.03 & 1.70 \\
& $p_1$ & - & --0.97 & --1.01 & --1.40 & --0.041 & --3.35 \\
& $q$ & - & 0.021 & 0.72 & 1.21 & 0.28 & --2.32 \\
& $R_{\mathrm{break}}$ & au & 49.89 & 43.43 & 48.19 & 61.95 & 74.77 \\
& $R_{\mathrm{out}}$ & au & 355.50 & 431.64 & 393.24 & 285.91 & 349.29 \\
\midrule
\multirow{3}{*}{Line width} & $L_{\mathrm{W}}$ & km\,s$^{-1}$ & 0.89 & 0.76 & 0.67 & 0.23 & 0.09 \\
& $p$ & - & --0.38 & --0.59 & --0.60 & --1.08 & --1.43 \\
& $q$ & - & 0.01 & 0.03 & 0.04 & --0.16 & --0.71 \\
\midrule
\multirow{2}{*}{Line slope} & $L_{\mathrm{S}}$ & - & 2.05 & 2.25 & 2.16 & 1.81 & 0.85 \\
& $p$ & - & 0.28 & 0.03 & 0.16 & 0.02 & 0.03 \\
\bottomrule
\end{tabular}
\tablefoot{\tablefoottext{a}{The inclination and position angle were fixed to the values obtained for the full $^{12}$CO cube.}}
\end{table}
\begin{table}[h!]
\small
\centering
\caption{Best-fit results of the modeling of the redshifted channels with the \textsc{discminer}\tablefootmark{(a)}.}\label{tab:fitsred}
\begin{tabular}{lccccccc}
\toprule
\toprule
Attribute &  Parameter & Unit & $^{12}$CO 7-6 & $^{12}$CO 3-2 & $^{12}$CO 2-1 & $^{13}$CO 2-1 & C$^{18}$O 2-1 \\
\midrule
\multirow{2}{*}{Orientation} & $x_{\mathrm{c}}$ & au & 0.13 & 4.33 & 2.11 & 0.28 & 1.33 \\
& $y_{\mathrm{c}}$ & au & 1.74 & --9.00 & --1.91 & --2.09 & --6.56 \\
\midrule
\multirow{2}{*}{Velocity} & $M_{*}$ & M$_{\odot}$ & 1.82 & 2.15 & 2.10 & 2.16 & 2.85 \\
& v$_{\mathrm{sys}}$ & km\,s$^{-1}$ & 5.41 & 5.57 & 5.62 & 5.62 & 5.65\\
\midrule
\multirow{4}{*}{Upper surface} & $z_{0}$ & au & 18.74 & 25.08 & 13.71 & 5.57 & 19.48 \\
& $p$ & - & 1.56 & 0.78 & 1.33 & 0.29 & 3.61 \\
& $q$ & - & 4.88 & 4.97 & 8.43 & 7.28 & 0.95 \\
& $R_{\mathrm{t}}$ & au & 276.79 & 334.32 & 319.07 & 234.89 & 45.80 \\
\midrule
\multirow{4}{*}{Lower surface} & $z_{0}$ & au & 12.66 & 40.26 & 14.18 & 2.52 & - \\
& $p$ & - & 1.72 & 1.92 & 2.11 & 0.23 & - \\
& $q$ & - & 2.18 & 0.66 & 1.14 & 3.07 & - \\
& $R_{\mathrm{t}}$ & au & 298.20 & 77.33 & 218.57 & 177.59 & - \\
\midrule
\multirow{6}{*}{Intensity} & $I_{0}$ & Jy\,px$^{-1}$ & 7.46 & 0.99 & 0.62 & 0.05 & 0.005 \\
& $p_0$ & - & --1.13 & 0.53 & --0.23 & 0.90 & 3.64 \\
& $p_1$ & - & --2.10 & --1.06 & --1.50 & --0.61 & --2.06 \\
& $q$ & - & 0.76 & 0.58 & 0.72 & 0.30 & -0.92 \\
& $R_{\mathrm{break}}$ & au & 61.38 & 43.16 & 46.73 & 61.69 & 96.24 \\
& $R_{\mathrm{out}}$ & au & 329.63 & 388.56 & 369.74 & 337.93 & 348.99 \\
\midrule
\multirow{3}{*}{Line width} & $L_{\mathrm{W}}$ & km\,s$^{-1}$ & 0.87 & 0.65 & 0.43 & 0.34 & 0.06 \\
& $p$ & - & --0.84 & --0.47 & --0.35 & --0.67 & --1.22 \\
& $q$ & - & 0.09 & --0.12 & --0.21 & --0.07 & --0.19 \\
\midrule
\multirow{2}{*}{Line slope} & $L_{\mathrm{S}}$ & - & 2.57 & 2.05 & 2.25 & 1.59 & 1.03 \\
& $p$ & - & 0.14 & 0.15 & 0.32 & 0.09 & --4.48 \\
\bottomrule
\end{tabular}
\tablefoot{\tablefoottext{a}{The inclination and position angle were fixed to the values obtained for the full $^{12}$CO cube.}}
\end{table}
\section{Comparison of the data and model channels}\label{appendix:chancomp}
\begin{figure*}[h!]
\centering
\includegraphics[width=1.0\textwidth]{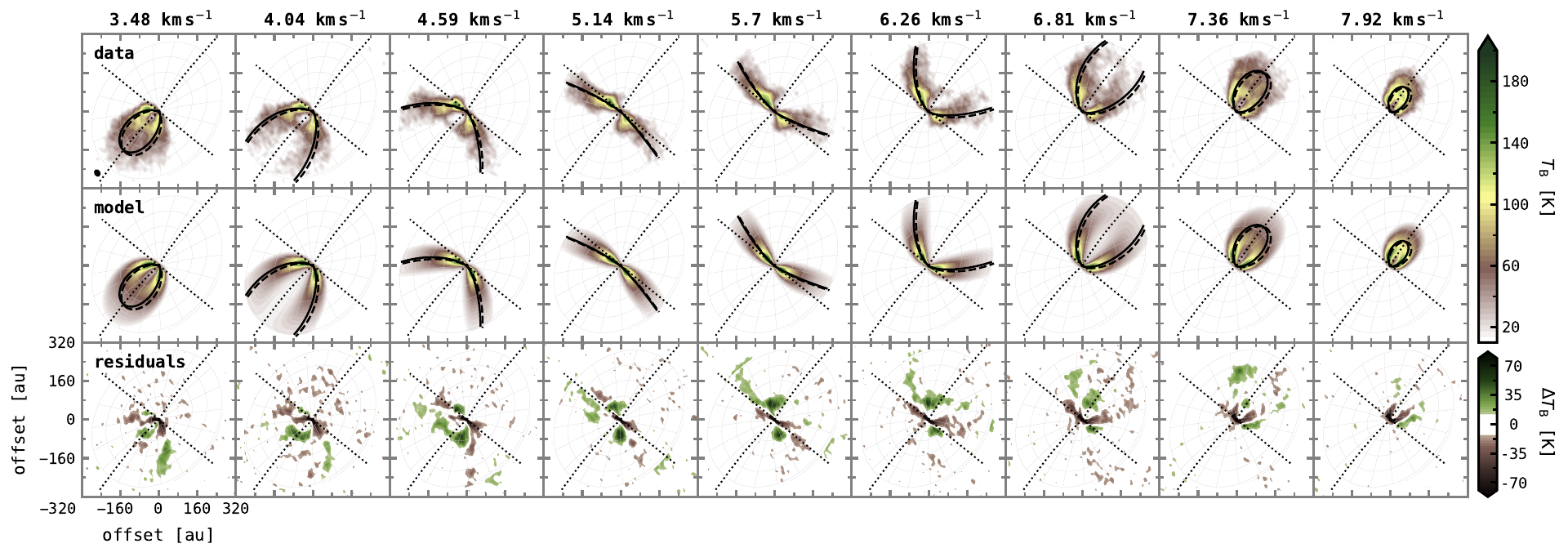}
\caption{Same as in \hyperref[fig:ChannelsComp12co21]{Fig.~\ref*{fig:ChannelsComp12co21}} but for $^{12}$CO 7-6.}\label{fig:ChannelsComp12co76}
\end{figure*}
\clearpage
\begin{figure*}[h!]
\centering
\includegraphics[width=1.0\textwidth]{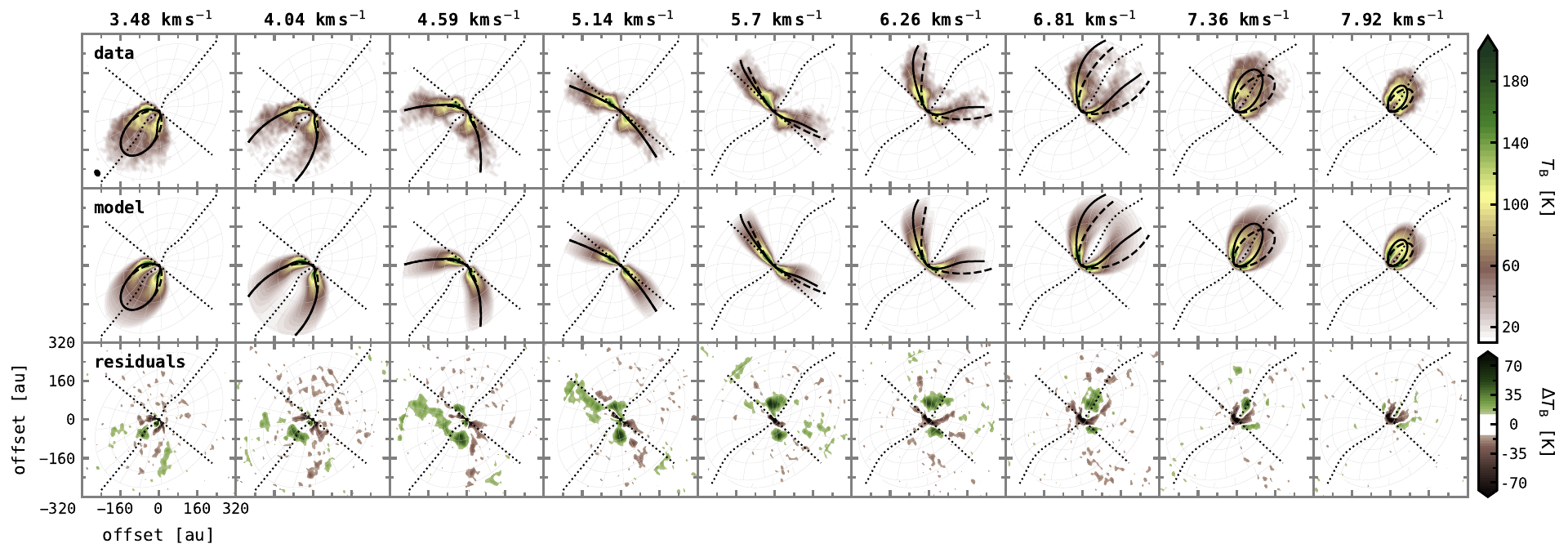}
\caption{Same as in \hyperref[fig:ChannelsComp12co21]{Fig.~\ref*{fig:ChannelsComp12co21}} but for $^{12}$CO 7-6 and the blue- and redshifted sides being modeled separately.}\label{fig:ChannelsComp12co76both}
\end{figure*}
\begin{figure*}[h!]
\centering
\includegraphics[width=1.0\textwidth]{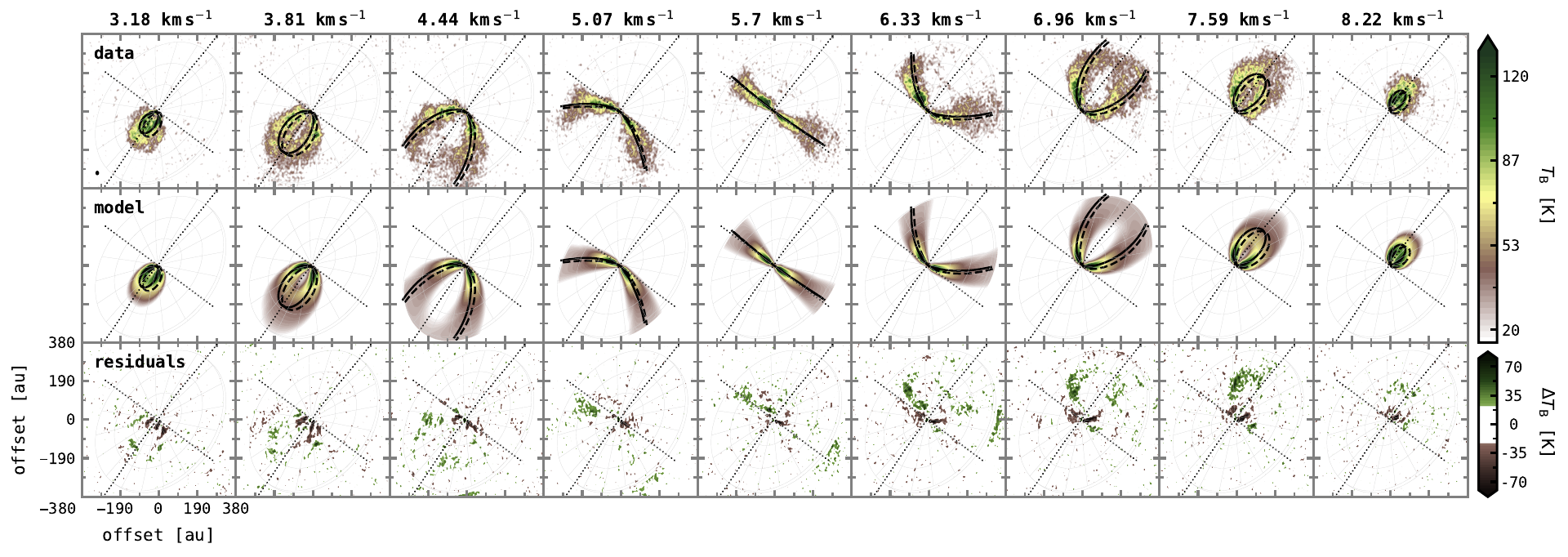}
\caption{Same as in \hyperref[fig:ChannelsComp12co21]{Fig.~\ref*{fig:ChannelsComp12co21}} but for $^{12}$CO 3-2.}\label{fig:ChannelsComp12co32}
\end{figure*}
%
%
\begin{figure*}[h!]
\centering
\includegraphics[width=1.0\textwidth]{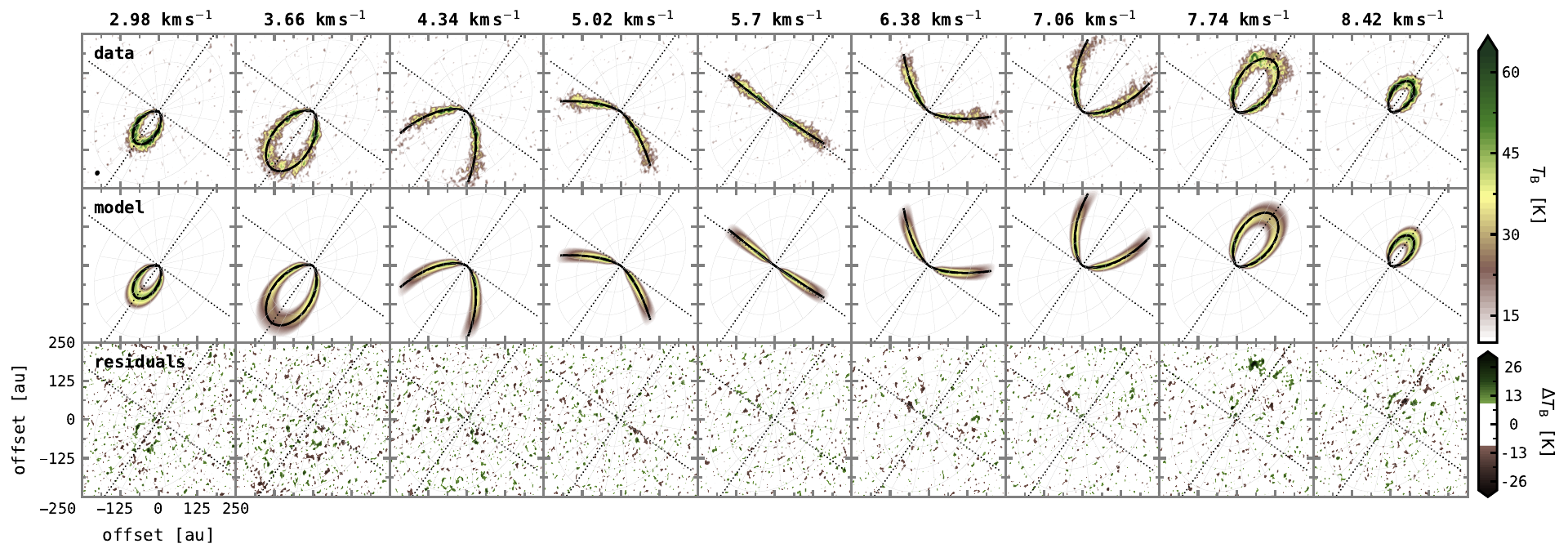}
\caption{Same as in \hyperref[fig:ChannelsComp12co21]{Fig.~\ref*{fig:ChannelsComp12co21}} but for $^{13}$CO 2-1.}\label{fig:ChannelsComp13co21}
\end{figure*}
\begin{figure*}[!h]
\centering
\includegraphics[width=1.0\textwidth]{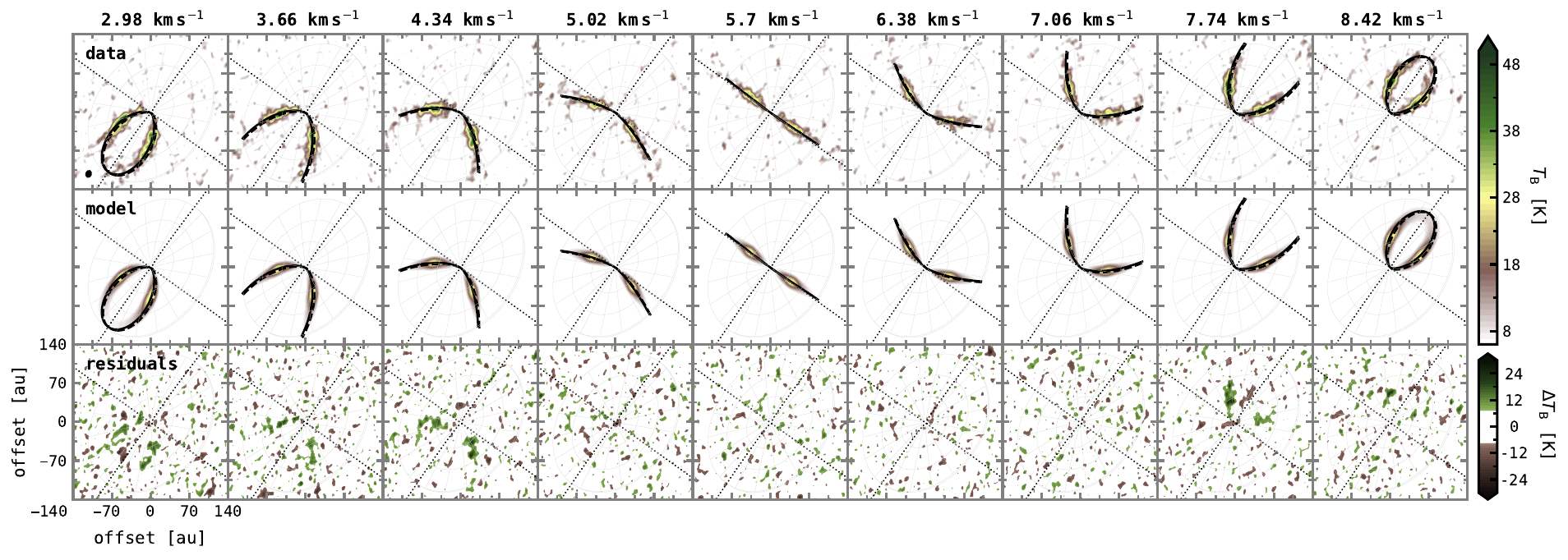}
\caption{Same as in \hyperref[fig:ChannelsComp12co21]{Fig.~\ref*{fig:ChannelsComp12co21}} but for C$^{18}$O 2-1.}\label{fig:ChannelsCompc18o21}
\end{figure*}
\section{Logarithmic spirals}
\begin{figure*}[h!]
\centering
\includegraphics[width=\textwidth]{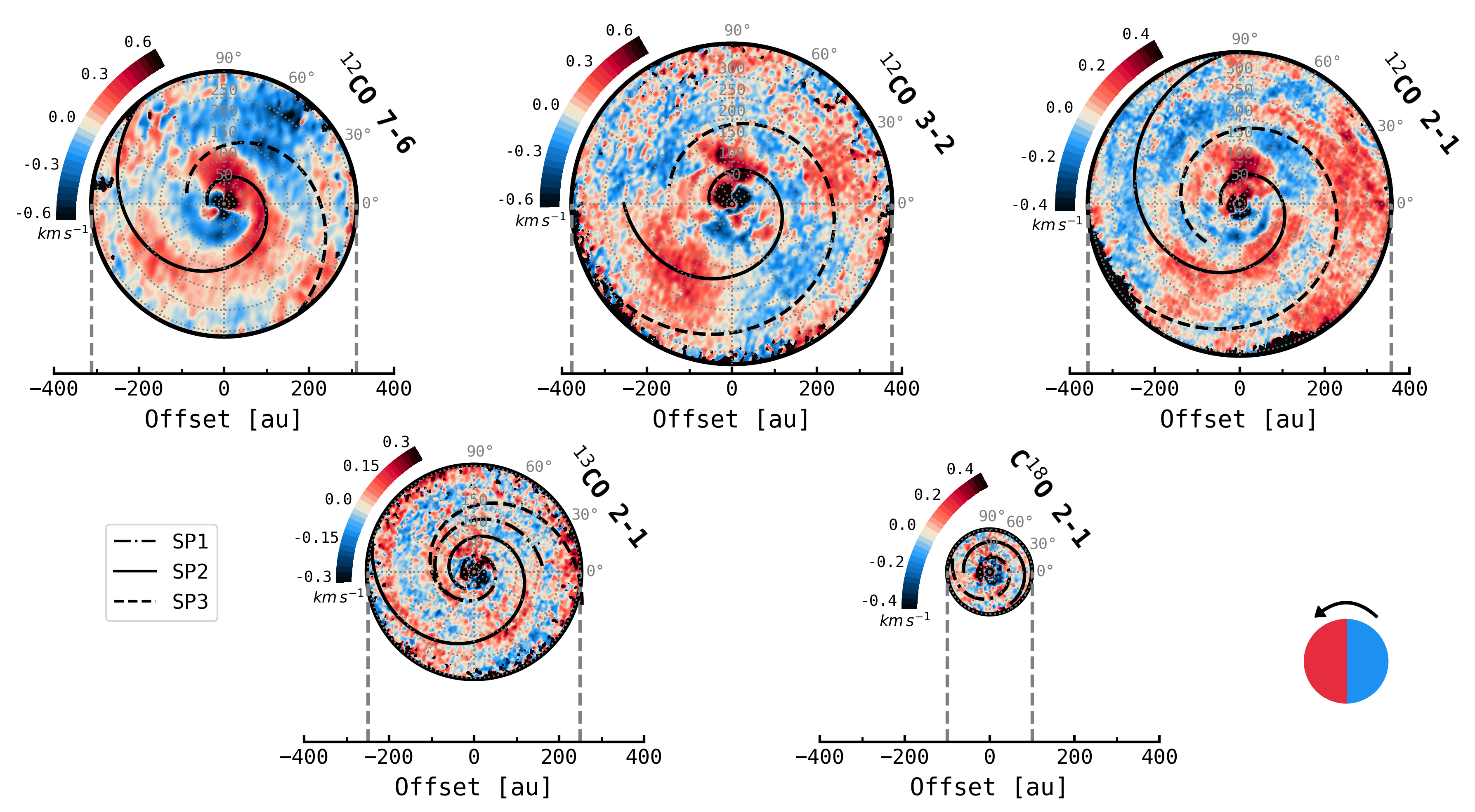}
\caption{Same as in \hyperref[fig:ResidualsVel]{Fig.~\ref*{fig:ResidualsVel}} but with overlaid logarithmic spirals.}\label{fig:ResidualsVelLog}
\end{figure*}
\clearpage
\section{Peak intensity and line width residuals}
\begin{figure*}[h!]
\centering
\includegraphics[width=\textwidth]{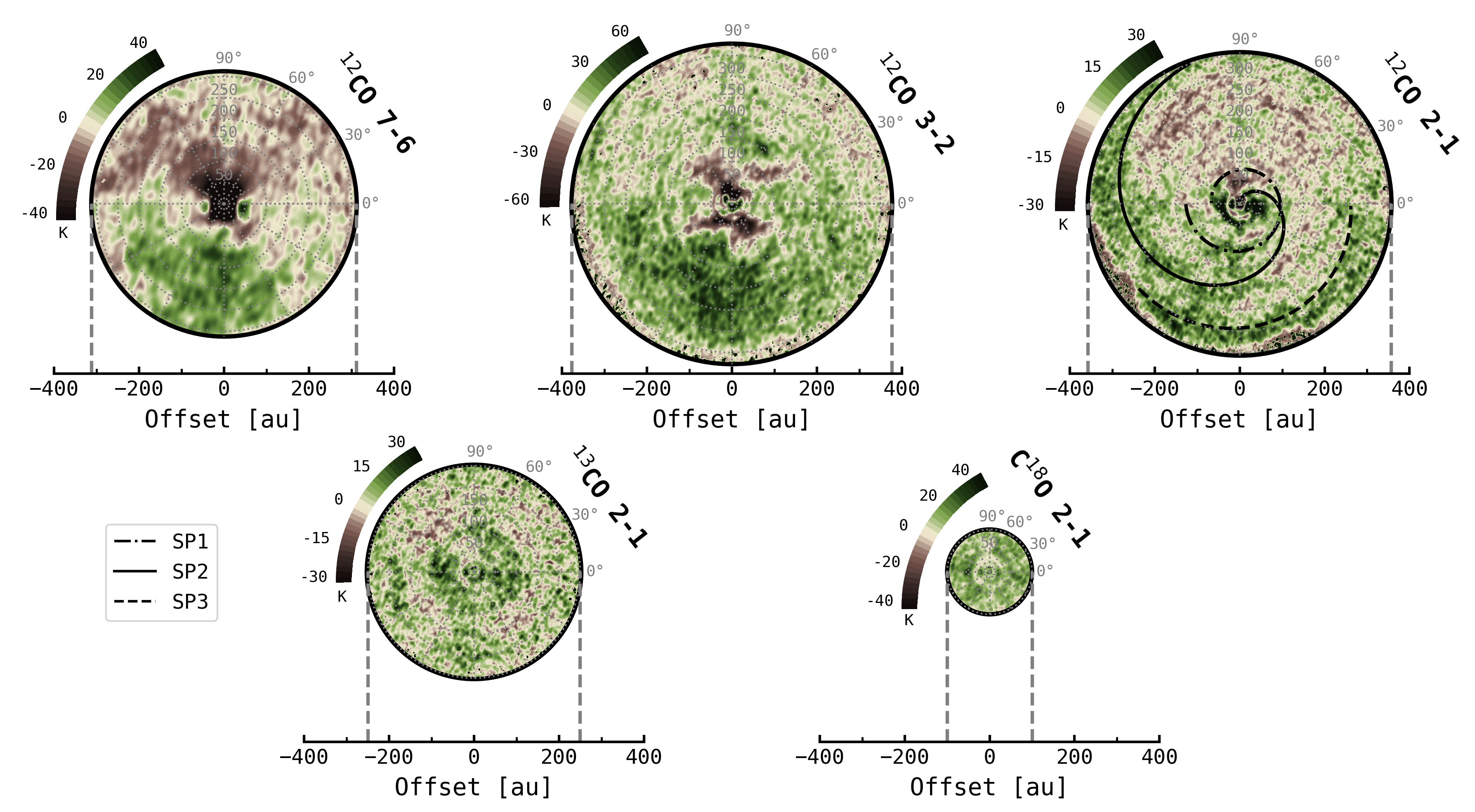}
\caption{Peak intensity residuals, shown for the five lines studied in this work.}\label{fig:ResidualsT}
\end{figure*}
\begin{figure*}[h!]
\centering
\includegraphics[width=\textwidth]{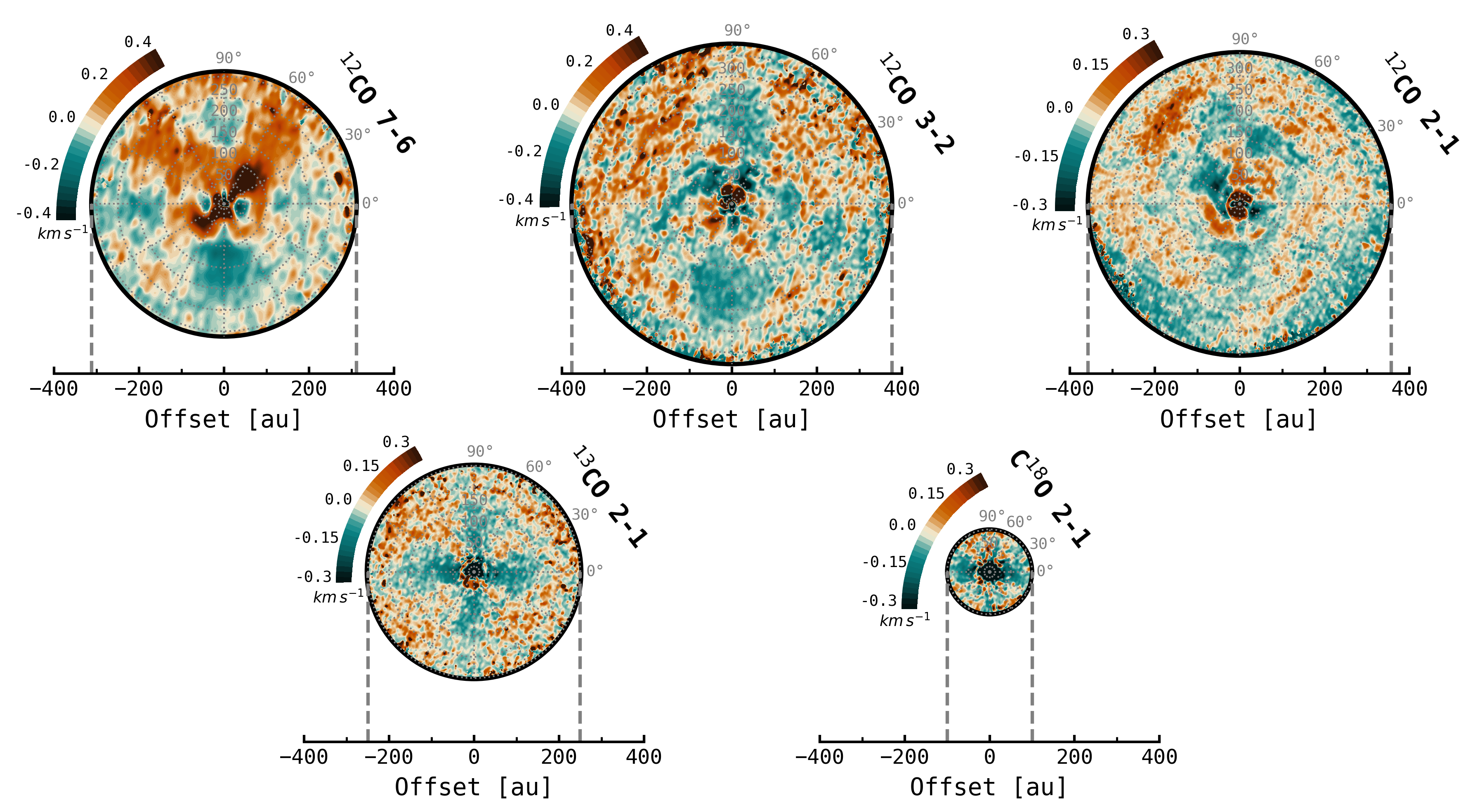}
\caption{Line width residuals, shown for the five lines studied in this work.}\label{fig:ResidualsLW}
\end{figure*}
\clearpage
\section{JvM-corrected data}\label{appendix:jvm}
In this section, we include some of the results for the JvM corrected data. This method \citep{Jorsater1995,Czekala2021} aims to correct for imaging artifacts induced by the non-Gaussianity of the dirty beam. The resulting epsilon values, the ratio of the CLEAN beam volume to the dirty beam volume, are given as 0.56 for Band 6 and 0.22 for Band 7. However, caution needs to be exercised when using the JvM correction \citep{Casassus2022b}, and thus we conducted our full analysis on both the corrected and non-corrected data sets. We find that our results are robust irrespective of the choice of using the JvM correction in the imaging process, with the difference that the non-corrected data have artificially lower noise levels by factors of 1/0.56 (Band6) and 1/0.22 (Band7), which affects mainly the patterns in the outer disk regions. The trends and substructures described in the main body of the paper are the same, yet the JvM-correction is appealing as it helps to smooth the data and make it easier to discern the patterns. Altogether, our conclusions are not affected by the use of the JvM-correction.
\begin{figure*}[h!]
\centering
\includegraphics[width=1.0\textwidth]{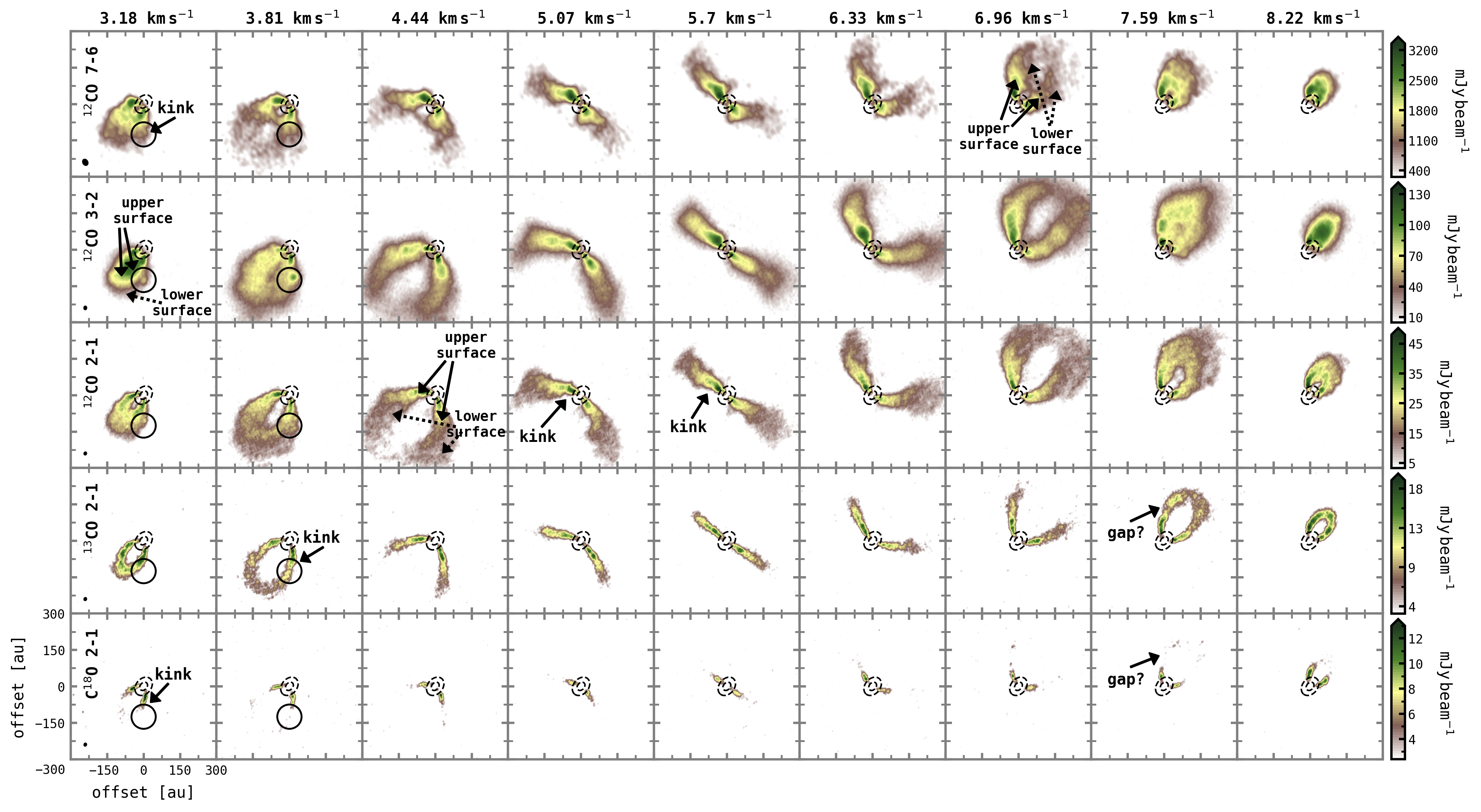}
\caption{Same as \hyperref[fig:Channels]{Fig.~\ref*{fig:Channels}} but for the JvM-corrected data.}\label{fig:Channels_jvm}
\end{figure*}
\begin{figure*}[h!]
\centering
\includegraphics[width=1\textwidth]{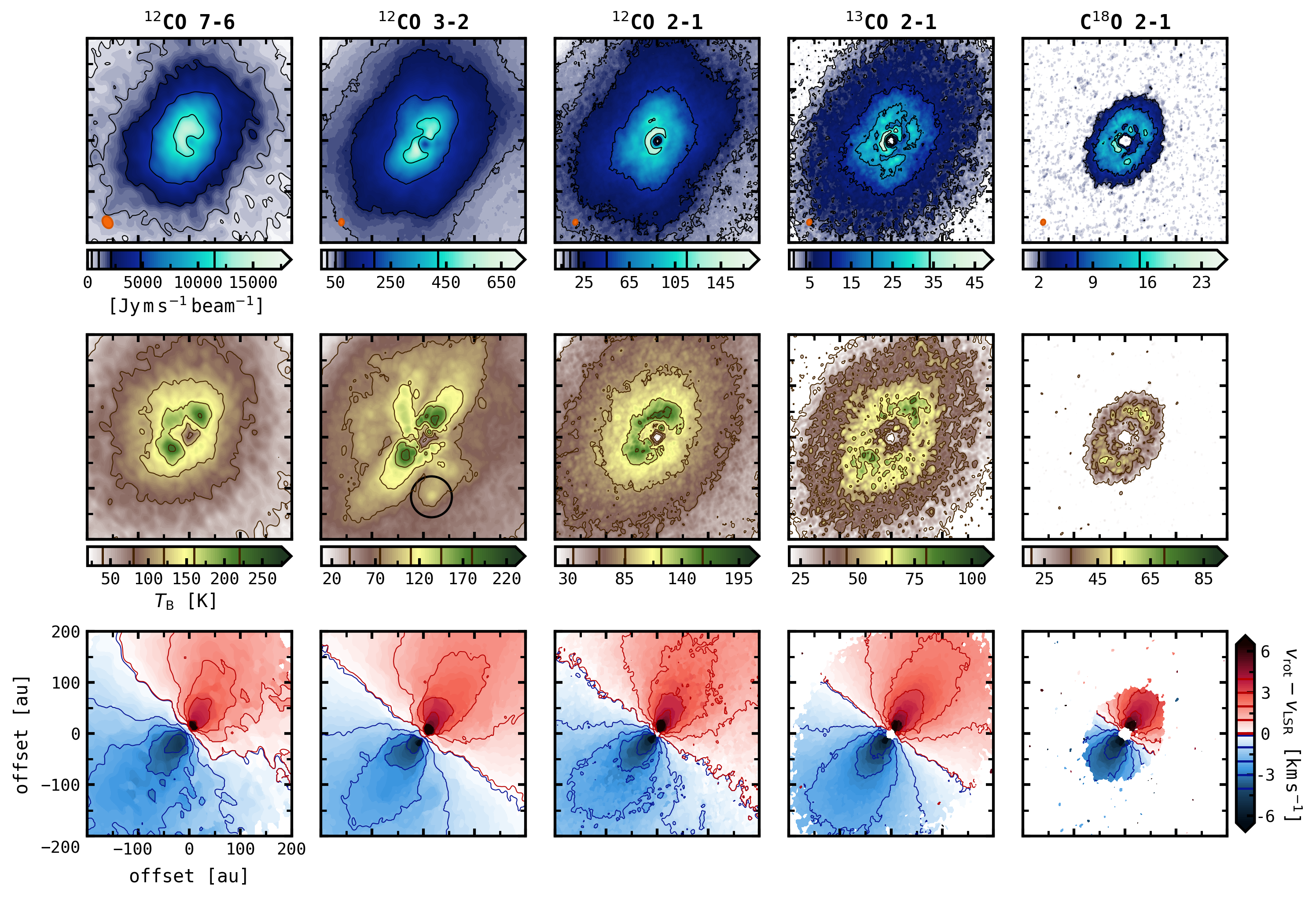}
\caption{Same as \hyperref[fig:Moments]{Fig.~\ref*{fig:Moments}} but for the JvM-corrected data.}\label{fig:Moments_jvm}
\end{figure*}
\begin{figure*}
\centering
\includegraphics[width=1.0\textwidth]{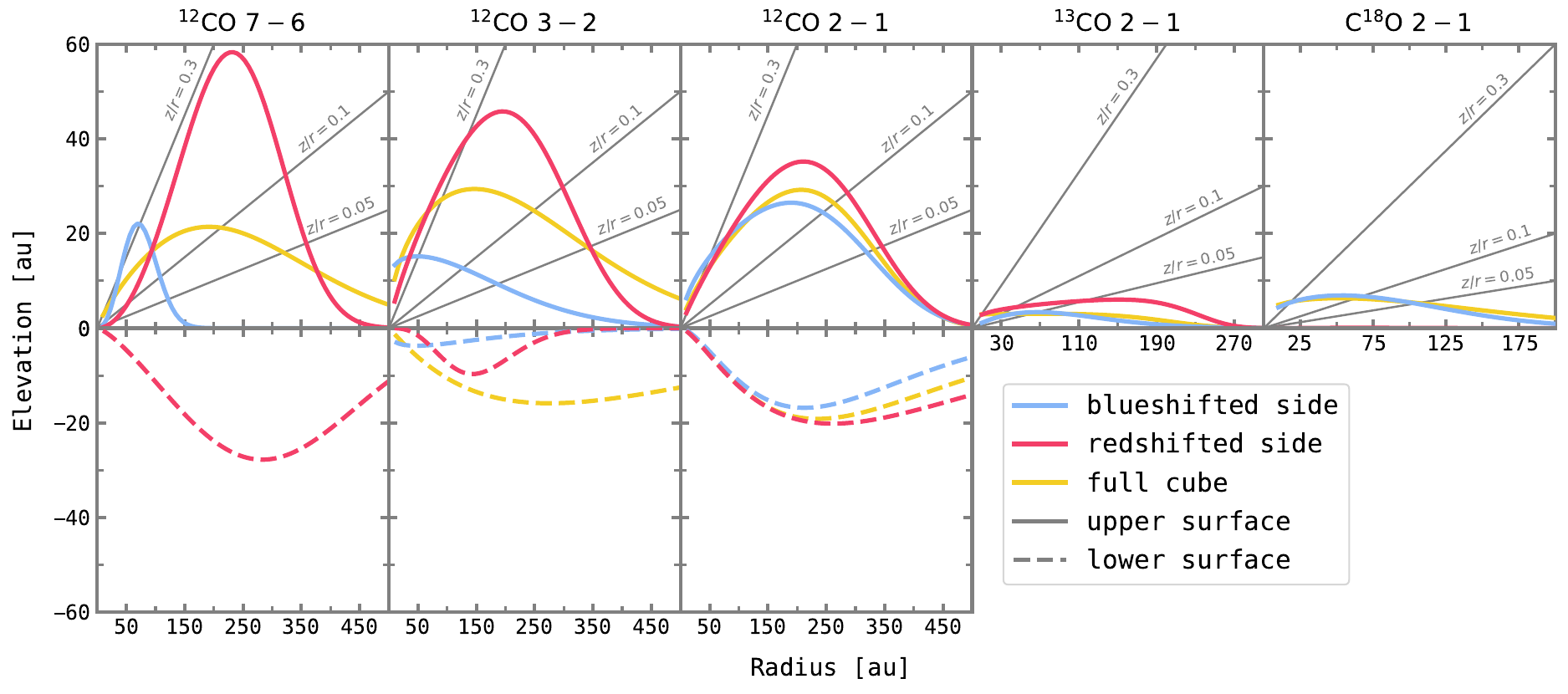}
\caption{Same as \hyperref[fig:Heights]{Fig.~\ref*{fig:Heights}} but for the JvM-corrected data.}\label{fig:Heights_nojvm}
\end{figure*}
\begin{figure*}
\centering
\includegraphics[width=1.0\textwidth]{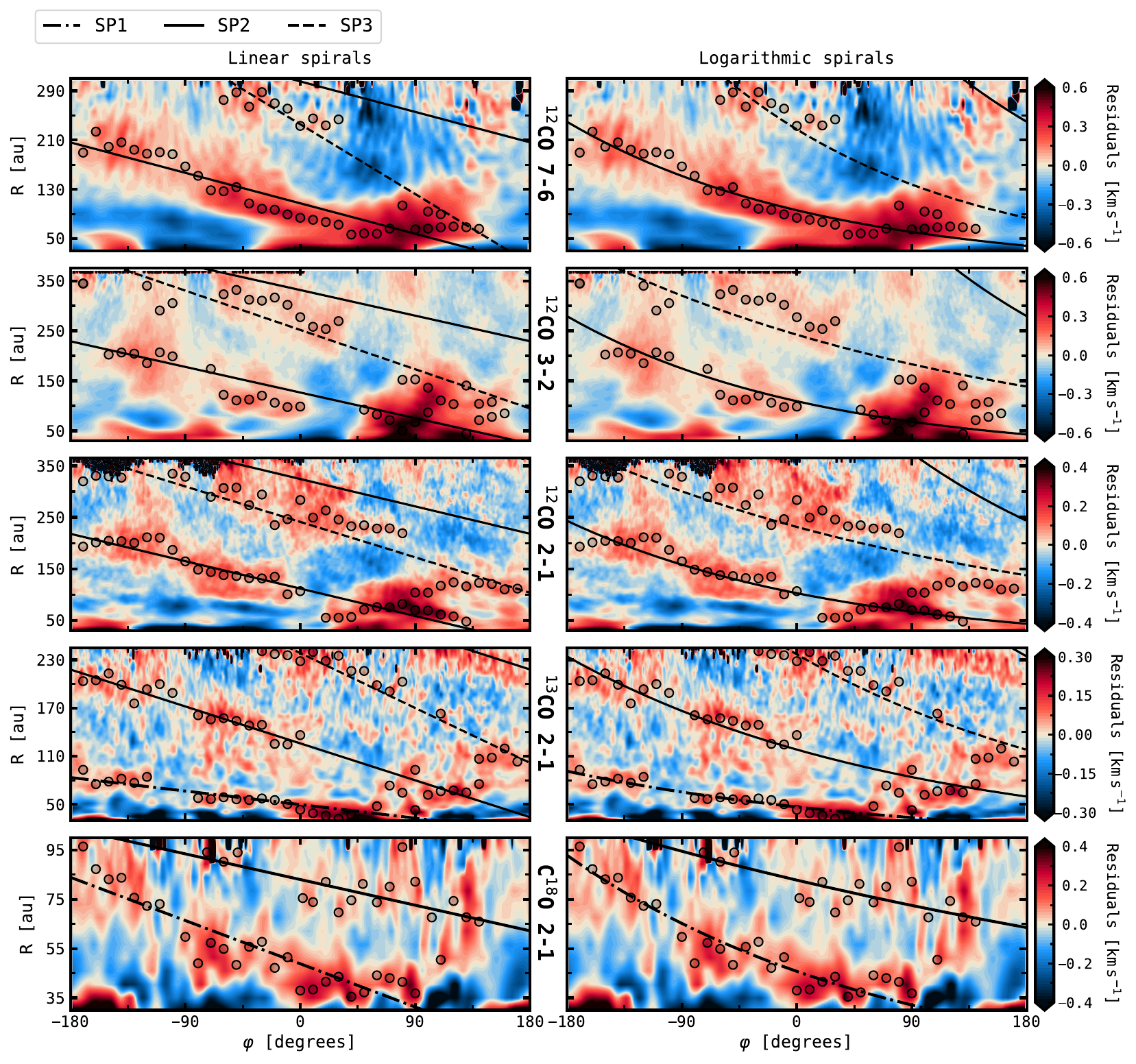}
\caption{Same as \hyperref[fig:PolarVel]{Fig.~\ref*{fig:PolarVel}} but for the JvM-corrected data.}\label{fig:PolarVel_jvm}
\end{figure*}
\clearpage
\twocolumn
\end{appendix}
\end{document}